\documentclass[11pt,
a4paper,
onecolumn,abstract=on]{scrartcl} 
\usepackage[american]{babel}	
\usepackage[autostyle=true]{csquotes}
\usepackage{microtype}  		
\usepackage{geometry}  			
\usepackage{graphicx}			
\usepackage{amsmath}			
\usepackage{bm}					
\usepackage{mathtools}			
\usepackage[ntheorem]{empheq}	
\usepackage{bbm}				
\usepackage{amsfonts} 			
\usepackage{yfonts}				
\usepackage[thref,amsmath,thmmarks,hyperref]{ntheorem}			
\usepackage[backend=biber,style=numeric-comp,giveninits=true,sorting=none]{biblatex}  
\usepackage[shortlabels]{enumitem}
\usepackage{accents}			
\usepackage{hyperref}			

\DeclareMathOperator{\Evalue}{E}
\DeclareMathOperator{\Var}{Var}
\DeclareMathOperator{\Cov}{Cov}

\DeclareMathOperator{\sgn}{sgn}

\newcommand{\abs}[1]{\lvert#1\rvert}	
\newcommand{\set}[2]{\left\lbrace #1 \middle\vert #2 \right\rbrace}
	
\renewcommand{\d}{\mathrm{d}}				
\renewcommand{\abs}[1]{\left\lvert #1 \right\rvert}

\newcommand{\R}{\mathbb{R}}				
\newcommand{\Z}{\mathbb{Z}}				
\newcommand{\N}{\mathbb{N}}				

\DeclareMathSymbol{\varnothing}{\mathord}{AMSb}{"3F}
\DeclareMathSymbol{\upharpoonright} {\mathrel}{AMSa}{"16}

\DeclareMathOperator{\pr}{pr}

\DeclareMathSymbol{\square}
{\mathord}{AMSa}{"03}
\DeclareMathSymbol{\blacksquare} {\mathord}{AMSa}{"04}

\renewcommand{\qedsymbol}{
								$\blacksquare$
							}
\newcommand{\oendmark}{
						$\diamondsuit$
						}

\theoremstyle{break}
\theoremheaderfont{\normalfont\bfseries}
\theorembodyfont{\normalfont 
				}
\theoremseparator{}
\theoremnumbering{arabic}
\theoremsymbol{\oendmark}
\newtheorem{Proposition}{Proposition}
\newtheorem{Theorem}
{Theorem}
\newtheorem{Lemma}
{Lemma}
\newtheorem{Corollary}
{Corollary}
{Conjecture}

\theoremsymbol{\oendmark}
\newtheorem{Definition}[Proposition]{Definition}

\theoremsymbol{\oendmark}
\newtheorem{Remark}
{Remark}
{Notation}

\newtheorem{Example}
{Example}

\theoremheaderfont{\normalfont\scshape}
\theorembodyfont{
		\normalfont
				}
\theoremstyle{nonumberplain}
\theoremseparator{}
\theoremsymbol{\qedsymbol} 
\newtheorem{Proof}{Proof}
\numberwithin{equation}{section}
\allowdisplaybreaks					
									
\renewbibmacro*{journal}{%
  \iffieldundef{shortjournal}
    {%
      \iffieldundef{journaltitle}
        {}
        {%
          \printtext[journaltitle]
            {%
              \printfield[titlecase]{journaltitle}%
              \setunit{\subtitlepunct}%
              \printfield[titlecase]{journalsubtitle}%
             }%
         }%
    }
    {\printtext[journaltitle]{\printfield{shortjournal}}}%
}

\AtEveryBibitem{\clearfield{month}}
\AtEveryBibitem{\clearfield{day}}
\AtEveryBibitem{\clearfield{number}}
\AtEveryBibitem{\clearfield{series}}
\AtEveryBibitem{\clearfield{eprintclass}}
\AtEveryBibitem{\clearfield{issn}}
\AtEveryBibitem{\clearfield{isbn}}

\title{%
		\vspace{- 2 \baselineskip}
		On the applicability of \\
		Kolmogorov’s theory of probability \\ 
		to the description of quantum phenomena. \\
		Part II: 
		Bell inequalities
		\vspace{-.25 \baselineskip}
		} 
 
\author{	%
			Maik Reddiger	\\
			\texttt{
			maik.reddiger@hs-anhalt.de
			}
			}

\date{September 25, 2026}

\begin{document}   

\maketitle 
\thispagestyle{empty}

\vspace{- 1.75 \baselineskip}

\begin{abstract} 
	\noindent
	Among the various ``no-go theorems'' in the foundations of quantum mechanics, 
	Bell inequalities are among the few, which do not implicitly rely on the mathematical 
	structure of quantum mechanics. Instead, 
	Bell inequalities are commonly understood to rest on 
	Kolmogorov's theory of probability 
	and a ``locality'' assumption. Physical evidence supporting their 
	violation is hence taken as evidence against 
	at least one of these two assumptions. 
	
	This work provides a comprehensive analysis of the physical debate on 
	Bell inequalities 
	from the point of view of Kolmogorov's theory of probability,  
	focusing on the 
	Bell-Clauser-Horn-Shimony-Holt inequality (BCHSH inequality). The central contribution  
	is the motivation and mathematical study of the notion of 
	contextuality in Bell experiments: in so-called contextual Bell models 
	the detectors themselves behave randomly 
	and the distribution of the random detector parameters depends on the chosen 
	detector settings. It is shown how even ``classical''  
	Bell experiments motivate such a description. 
	The need to account for contextuality was pointed out before by 
	de la Pe\~na, Cetto, Brody, Lochak, Bohm, Hiley and others. 
	A broad class of contextual models is proven to satisfy the BCHSH inequality, thereby 
	generalizing a previous result by Gill and Lambare. Nonetheless, 
	even under strong locality requirements, contextual 
	Bell models may violate the BCHSH inequality: 
	based on work by Popescu and Rohrlich, we consider  
	a ``proof of concept'' example exhibiting the maximum possible violation. 
	
	Contrary to the much debated
	``loopholes'' in empirical tests of Bell inequalities, contextuality thus 
	reveals a conceptual misunderstanding in the physical conclusions commonly 
	drawn from Bell inequality violations. 
	
	This work is mostly self-contained. 
	Part I of this series critically reflects on the relationship between quantum 
	mechanics and Kolmogorov's theory of probability, showing how the 
	former may be fruitfully adapted to comply with the latter.
	Part III addresses the projection 
	postulate and the question of measurement.  
\end{abstract}	

\vspace{.25 \baselineskip}

{
	Contextuality
	\ - \ 
	Quantum nonlocality
		\\[.5 \baselineskip]
	\\	
}
\pagestyle{headings}    

\tableofcontents

\section{Introduction}
\label{sec:intro}

In the foundations of 
quantum theory, ``no-go theorems'' aim to prove the incompatibility of quantum 
physics with so-called ``hidden variables'': descriptions of quantum systems, 
which quantum mechanics alone cannot account for 
\cite{belinfanteSurveyHiddenVariablesTheories1973,
homeEnsembleInterpretationsQuantum1992}. 

The first mathematical statement against such ``hidden parameters'' was provided by 
von Neumann in 1932 
\cite{vonneumannMathematischeGrundlagenQuantenmechanik1932,
vonneumannMathematicalFoundationsQuantum1955}. 
Von Neumann was well aware of how statistical mechanics had 
reduced a sizable part of the foundations of thermodynamics to those of 
Newtonian mechanics. The concern that the laws of 
quantum mechanics could similarly be reduced to those of a more fundamental 
theory weighed heavily upon him 
\cite{vonneumannMathematischeGrundlagenQuantenmechanik1932,
vonneumannMathematicalFoundationsQuantum1955}:  
\begin{quote}
	Whether or not an explanation of this type, by means of hidden
	parameters, is possible for quantum mechanics, is a much discussed
	question. The view that it will sometime be answered in the affirmative
	has at present prominent representatives. 
	If it were correct, it would
	brand the present form of the theory as provisional, since then the
	description of the states would be essentially incomplete.
\end{quote} 
The threat von Neumann attempted to tame with his 
``no-go theorem'' was thus the possibility of a more refined description 
of quantum phenomena than what quantum mechanics is able to provide. 
Though his argument was later shown to be baseless 
\cite{hermannNaturphilosophischenGrundlagenQuantenmechanik1935,
bellProblemHiddenVariables1966,bellSpeakableUnspeakableQuantum2004}, 
as a political intervention it was highly successful: for 
over thirty years it had served to discredit approaches aiming for such a completion 
(cf. 
Sec. 4 in \cite{pinchWhatDoesProof1977} as well as 
\cite{belinfanteSurveyHiddenVariablesTheories1973,
pinchHiddenvariablesControversyQuantum1979, 
freirejuniorQuantumDissidentsRebuilding2015,budroniKochenSpeckerContextuality2022,
baggottTheoremJohnBell2024}). The opposition 
spearheaded by Einstein (cf. Sec. 7.1 in \cite{bacciagaluppiQuantumTheoryCrossroads2009} and 
\cite{einsteinCanQuantumMechanicalDescription1935}) 
had been defeated in the political arena. 

In the 1950s Bohm's constitutive works 
\cite{bohmSuggestedInterpretationQuantum1952,bohmSuggestedInterpretationQuantum1952a} 
provided an incentive for a reexamination of von Neumann's result 
\cite{bellProblemHiddenVariables1966,bellSpeakableUnspeakableQuantum2004,
pinchWhatDoesProof1977,freireStoryEndingQuantum2003,
freirejuniorQuantumDissidentsRebuilding2015}. The genre of ``no go theorems'' 
was revitalized by the works of 
Gleason \cite{gleasonMeasuresClosedSubspaces1957} in 1957 
as well as Jauch and Piron \cite{jauchCanHiddenVariables1963} in 1963. In 1975 Kochen 
and Specker proved the theorem bearing their name 
\cite{kochenProblemHiddenVariables1975}. In 1990 Peres and Mermin 
\cite{peresIncompatibleResultsQuantum1990,merminSimpleUnifiedForm1990,peresTwoSimpleProofs1991,
merminHiddenVariablesTwo1993} provided a similar, more illustrative result in 
four dimensions, 
the ``Peres-Mermin magic square'' \cite{larssonContextualExtensionSpekkens2012,
krishnaDerivingRobustNoncontextuality2017}. 
For a modern exposition of 
these statements, interested readers are referred to Chap. 5 in   
\cite{morettiFundamentalMathematicalStructures2019}. 

It is a common thread in all of the aforementioned ``no go theorems'' that they assert 
the validity of certain aspects of quantum probability theory and then show 
the incompatibility with ``classical'' notions of probability. In other words, 
they may be viewed as results on the incompatibility of 
quantum probability theory 
\cite{accardiTopicsQuantumProbability1981,accardiFoundationsQuantumProbability1982,
streaterClassicalQuantumProbability2000,redeiQuantumProbabilityTheory2007} 
with Kolmogorov's theory of probability 
\cite{kolmogoroffGrundbegriffeWahrscheinlichkeitsrechnung1933,
kolmogorovFoundationsTheoryProbability1956,klenkeProbabilityTheoryComprehensive2020}. 
Nowadays and outside of quantum physics, the latter is acknowledged as the 
standard theory of probability. It was axiomatized by Kolmogorov 
in 1933 \cite{kolmogoroffGrundbegriffeWahrscheinlichkeitsrechnung1933,
kolmogorovFoundationsTheoryProbability1956,
reddigerAddendumApplicabilityKolmogorovs2026}, and in the foundations of quantum 
theory it tends to be indiscriminately equated with ``hidden variables'' 
(cf. \cite{szaboRealMeaningBells1994}). From here on, we shall refer to Kolmogorov's theory of probability as mathematical probability theory. 

In his 1966 article \cite{bellProblemHiddenVariables1966,
bellSpeakableUnspeakableQuantum2004}, 
Bell pointed out that such ``no go theorems'' 
rely on an implicit circular argument and are thus not empirical in nature: 
it is asserted that only quantum mechanics 
can provide an adequate description of quantum phenomena. It is assumed 
what is supposedly shown. 
In Bell's first main 
theorem \cite{bellEinsteinPodolskyRosen1964,bellSpeakableUnspeakableQuantum2004}, 
in contrast, he asserted the validity of mathematical 
probability theory and showed that under certain additional 
assumptions a general inequality would hold, which violates a specific 
quantum-mechanical prediction. 
This so-called Bell inequality and subsequent ones 
\cite{clauserProposedExperimentTest1969,
bellIntroductionHiddenvariableQuestion1971,bellSpeakableUnspeakableQuantum2004,
clauserExperimentalConsequencesObjective1974} 
thus turned the question of the applicability of mathematical 
probability theory to quantum phenomena into one, which was at least in principle 
subject to empirical scrutiny. 

Nowadays, the list of such Bell (test) experiments documented in 
the literature is long, see e.g. 
\cite{freedmanExperimentalTestLocal1972,
holtAtomicCascadeExperiments1973,
clauserExperimentalInvestigationPolarization1976,
fryExperimentalTestLocal1976,
aspectExperimentalTestsRealistic1981,
aspectExperimentalRealizationEinsteinPodolskyRosenBohm1982,
aspectExperimentalTestBell1982,
ouViolationBellsInequality1988,
shihNewTypeEinsteinPodolskyRosenBohm1988,
rarityExperimentalViolationBells1990,
kiessEinsteinPodolskyRosenBohmExperimentUsing1993,
kwiatNewHighIntensitySource1995,
tittelViolationBellInequalities1998,
weihsViolationBellInequality1998,
panExperimentalTestQuantum2000,
roweExperimentalViolationBell2001,
matsukevichEntanglementRemoteAtomic2006,
ursinEntanglementbasedQuantumCommunication2007,
matsukevichBellInequalityViolation2008,
ansmannViolationBellsInequality2009,
hofmannHeraldedEntanglementWidely2012,
vermeydenExperimentalViolationThree2013,
giustinaBellViolationUsing2013,
christensenDetectionLoopholeFreeTestQuantum2013,
hensenLoopholefreeBellInequality2015,
giustinaSignificantLoopholeFreeTestBell2015,
shalmStrongLoopholeFreeTest2015,
rosenfeldEventReadyBellTest2017,
yinSatellitebasedEntanglementDistribution2017}. The general 
consensus is that 
Bell inequalities are indeed empirically 
violated under appropriate conditions. 

Nonetheless, the physical interpretation of these empirical results  
is in essence a matter of theory. 

The most common conclusion drawn is that 
``local realism'' -- that is, a description via mathematical probability theory 
and suitable requirements of ``locality'' -- is not a physically tenable position 
\cite{brunnerBellNonlocality2014,brunnerPublishersNoteBell2014,aspectClosingDoorEinstein2015,
freirejuniorAlainAspectExperiments2022,
nobelcommitteeforphysicsScientificBackgroundNobel2022}. Other, arguably 
more radical interpretations have also been suggested (see e.g.  
\cite{vervoortBellTheoremTwo2013,
whartonQuantumStatesOrdinary2014,whartonColloquiumBellTheorem2020,
hossenfelderRethinkingSuperdeterminism2020,
vaidmanBellInequalityManyWorlds2016,
brownBellBellsTheorem2016}). 

The purpose of this work is to provide a comprehensive theoretical analysis of the physical 
debate on Bell inequalities on the basis of the following hypotheses: 
\begin{enumerate}[1)]
	\item	Mathematical probability theory provides a valid description 
			of quantum phenomena in general 
			and of Bell experiments in particular, and 
	\item 	there are Bell experiments, in which the Bell inequalities discussed here 
			are violated. 
\end{enumerate}

While this work provides an in-depth analysis of ``locality'' in Bell 
experiments, the commonly stated conclusion that under the above 
hypotheses ``nonlocality'' 
must 
follow 
\cite{bellLocalityQuantumMechanics1975,
durrQuantumPhysicsQuantum2013,
brunnerBellNonlocality2014,
brunnerPublishersNoteBell2014,
dewdneySpinNonlocalityQuantum1988,
bohmNonlocalityLocalityStochastic1989,
beyerSternGerlachEPRB2024,vaidmanEinsteinWasNot2026,lucWhatAreBearers2026,
dartoisBellExperimentsRevisited2026} 
is not supported. 

This conclusion follows from the central result of this work: a third 
major assumption 
in the derivation of Bell inequalities has been neglected in the physical 
debate---noncontextuality. As we shall see, 
even in ``classical'' Bell experiments noncontextuality is a stark assumption. 
And without this assumption, Bell inequalities may be violated while  
suitable notions of locality are respected. 

The concept of contextuality first entered the debate  
in Shimony's 1971 article  
\cite{shimonyExperimentalTestsLocal1971} (cf. Appx. A in \cite{budroniKochenSpeckerContextuality2022}).  
In the later literature, 
this came to be known as ``Kochen-Specker contextuality'' 
(cf. \cite{kochenProblemHiddenVariables1975,budroniKochenSpeckerContextuality2022} 
and also \cite{spekkensContextualityPreparationsTransformations2005,
adlamFoundationsQuantumMechanics2021,grangierKolmogorovianCensorshipPredictive2026}). 
In this work, however, contextuality carries a slightly different meaning: 
it is based on Kolmogorov's assertion that an application of the 
theory of probability ``to the actual world of experiments'' requires a 
``complex of conditions''---an experimental context, that is  
(cf. Sec. I.2 in \cite{kolmogoroffGrundbegriffeWahrscheinlichkeitsrechnung1933,
kolmogorovFoundationsTheoryProbability1956} as well as 
\cite{khrennikovLocalRealismContextualism2002,
khrennikovContextualApproachQuantum2009,
khrennikovProbabilityRandomnessQuantum2016,
khrennikovContextualReinterpretationQuantum2025}  
and the philosophical works 
\cite{vonmisesWahrscheinlichkeitsrechnung1931,vonmisesProbabilityStatisticsTruth1957}). 
Different experimental contexts thus correspond to different random experiments and, 
a priori, require different probability spaces for their description. 
Contextuality, as understood here, is therefore rooted in the mathematical theory of 
probability, while ``Kochen-Specker contextuality'' is rooted in quantum mechanics. The 
close relationship between the two concepts nonetheless justifies the use of the 
same word and at times both notions have been used interchangeably in the 
quantum theory literature (cf. 
\cite{genoveseResearchHiddenVariable2005,terracunhaMeasuresMeasurementsFibre2019,
garolaKolmogorovianNonKolmogorovianProbabilities2021,
tezzinViolatingKCBSInequality2025,
papatryfonosProposedExperimentsDetecting2025,
skottDistinguishingBohmianContextuality2026}). 

Contextuality is potentially relevant in Bell experiments, because Bell inequalities 
require different detector settings for their empirical verification. It is therefore a 
nontrivial assumption to require the probability measure describing the experiment 
to be independent of the detector settings. 
This limitation in the derivation of Bell inequalities was pointed out before by 
de la Pe\~na, Cetto and Brody \cite{delapenaHiddenvariableTheoriesBell1972}, 
Lochak \cite{lochakHasBellsInequality1976}, as well as 
Bohm and Hiley \cite{bohmNonlocalityQuantumTheory1981}. 
Since then, other authors raised similar objections  
\cite{pitowskyResolutionEinsteinPodolskyRosenBell1982,
demuynckBellInequalitiesTheir1986,bransBellsTheoremDoes1988,
szaboRealMeaningBells1994,
feldmannNewLoopholeEinsteinPodolskyRosen1995,
nagasawaLocalityHiddenvariableTheories1997,
khrennikovContextualistViewpointGreenberger2001,
khrennikovMathematicianViewpointBell2007,
khrennikovViolationBellsInequality2009,
nieuwenhuizenWhereBellWent2009,
khrennikovProbabilityRandomnessQuantum2016,
kupczynskiCanWeClose2017,
kupczynskiEinsteinianNosignallingViolated2017,
hanceBellsTheoremAllows2022,
kupczynskiContextualityNonlocalityWhat2023,
papatryfonosStaticBellTest2024,
arroyoFamilyDeterministicModels2025,
papatryfonosProposedExperimentsDetecting2025,
lucWhatAreBearers2026,
bacciagaluppiExtendingBellsTheorem2026}. Bell's 
1975 article \cite{bellLocalityQuantumMechanics1975,bellSpeakableUnspeakableQuantum2004} 
shows that he was vaguely aware of this argument, yet he discarded 
it on the basis that the ``hidden variables'' could solely 
correspond to an array of ``initial values'' and that their distribution would 
therefore be independent of the detector settings (see also 
Sec. 5 in \cite{bellBertlmannsSocksNature1981} or 16.5 in 
\cite{bellSpeakableUnspeakableQuantum2004}). This assertion is a fallacy, 
as may be seen in an illustrative physical example, 
Exs. \ref{Ex:correlatedP} and  \ref{Ex:correlatedP_cont}. A more 
in-depth discussion of the role of dynamics in Bell experiments is given 
in Sec. \ref{sec:time}). 

The analysis provided here is motivated by the specific question of 
what the empirical violation of Bell inequalities implies for 
\emph{geometric quantum theory} 
\cite{reddigerMadelungPictureFoundation2017,
reddigerProbabilisticFoundationNonrelativistic2022,
reddigerApplicabilityKolmogorovsTheory2025,reddigerAddendumApplicabilityKolmogorovs2026}. 
Geometric quantum theory is a natural adaption of quantum mechanics as well as 
relativistic quantum theory 
\cite{reddigerProbabilisticFoundationRelativistic2024} to mathematical 
probability theory. In part I of this series 
\cite{reddigerApplicabilityKolmogorovsTheory2025,reddigerAddendumApplicabilityKolmogorovs2026}, a mathematically 
rigorous theory for non-relativistic $N$-body quantum systems subject to a time-independent 
scalar potential was constructed. It was shown that the theory reproduces central 
predictions of quantum mechanics for such systems but that in certain cases it 
is nonetheless capable of 
making distinct predictions. The present work is one in a three-part series 
in support of the hypothesis that mathematical probability theory  
applies to the description of quantum phenomena. Indeed, quantum physics is 
the last remaining domain of empirical science, where mathematical probability theory  
has not been (fully) co-opted. As argued above, among the various 
``no-go theorems'' those that evolved out of Bell's theorem 
\cite{bellProblemHiddenVariables1966,bellSpeakableUnspeakableQuantum2004} 
are essentially the only ones of empirical 
relevance. 

This work is mostly self-contained, so part I 
\cite{reddigerApplicabilityKolmogorovsTheory2025,reddigerAddendumApplicabilityKolmogorovs2026} is not a prerequisite to 
comprehension. In Sec. \ref{sec:review} we review three notable theorems 
in the debate on Bell inequalities. There our subsequent focus on the 
Bell-Clauser-Horn-Shimony-Holt inequality (BCHSH inequality) is justified. In 
Sec. \ref{sec:discussion} we abstract the concept of a Bell model from these theorems 
and study corresponding notions of locality. The more general concept of a contextual 
Bell model is motivated, introduced, and studied in Sec. \ref{sec:generalizedBell}. 
There we consider a number of ``classical'' Bell experiments, which call for such 
a description. Still, we also find that some 
contextual Bell models can be reduced to the ``noncontextual'' case and that, moreover,  
a broad class of contextual Bell models do satisfy the BCHSH inequality. 
While this does make the role of contextuality in Bell experiments more subtle 
(cf. Rem. \ref{Rem:contextuality}), 
Sec. \ref{sec:poc_example} provides an explicit example that even under a strong 
locality constraint contextuality can lead to 
a violation of the BCHSH inequality. 
Indeed, the example employs results by Popescu and Rohrlich 
\cite{popescuQuantumNonlocalityAxiom1994} to attain the maximum possible 
violation above the Tsirel'son bound 
(cf. \cite{cirelsonQuantumGeneralizationsBells1980} and p. 174 in 
\cite{peresQuantumTheoryConcepts2002}). 
Sec. \ref{sec:time} provides a condensed, more philosophical account of 
time evolution in Bell experiments, aiming to provide a physical argument of 
how it can lead to a violation of the BCHSH 
inequality in the presence of only ``local interactions''.  
The argument is related to, yet nonetheless distinct from the one by 
Hess and Philipp 
\cite{hessEinsteinseparabilityTimeRelated2001,
hessPossibleLoopholeTheorem2001,hessExclusionTimeTheorem2002}. 
The connection to statistical 
data for general contextual Bell models (including ``noncontextual'' ones) is 
drawn in Sec. \ref{sec:statistics}. Sec. \ref{sec:conclusion} concludes 
the discussion and provides an outlook for further research. 

Readers only interested in how contextuality 
enables a violation of 
the BCHSH inequality without violating ``locality'' are advised to 
skim through Sec. \ref{sec:review}, paying particular attention to the definition of 
a Bell experiment (cf. Fig. \ref{fig:Bellexperiment}), 
to then jump to Defs. \ref{Def:Bellmodel} as well as \ref{Def:locality} and afterwards 
read Sec. \ref{sec:generalizedBell} in some detail. This should suffice 
to understand the example in Sec. \ref{sec:poc_example}. 

Even though this work is fundamentally concerned with quantum theory, 
readers are expected to be familiar with basic concepts in 
probability theory: measurable spaces, probability spaces, random variables, 
expectation values, 
joint and marginal distributions, conditional probabilities, 
the phrase ``almost surely'', stochastic dependence and correlation. 
There are many textbooks on such 
topics; here we shall frequently refer to the one by Klenke 
\cite{klenkeProbabilityTheoryComprehensive2020}.

With regards to notation, we shall use standard conventions from probability theory. 
The author believes that problematic notation has clouded many arguments 
in this subject area, so this is a conscious choice. 
In particular, if  
$X \colon \Lambda \to \R$ is a random variable on the probability space $(\Lambda, \mathcal{A},\mathbb{P})$, then for any $A \in \mathcal{A}$ we shall write 
\begin{equation}
	\mathbb{P}(X \in A ) 
	= \mathbb{P} \left(\set{\lambda \in \Lambda}{X(\lambda)  \in A }\right)
	= \mathbb{P} \left(X^{-1}(A)\right) = 
	\left( \mathbb{P} \circ X^{-1} \right)(A)	
	 \ .
\end{equation}
Thus, $\mathbb{P} \circ X^{-1}$ denotes the distribution of $X$. If $A = \lbrace x 
\rbrace$ for some $x \in \R$, then we write 
\begin{equation}
	\mathbb{P}(X = x ) 
	= \mathbb{P} \left(\set{\lambda \in \Lambda}{X(\lambda) = x}\right)
	= \left( \mathbb{P} \circ X^{-1} \right)(\lbrace x \rbrace ) \ . 
\end{equation}
In general, we use $\mathbb{E}(X)$ to denote the expectation value of 
$X$, if it exists. $\mathcal{B}(U)$ denotes the Borel sets of 
some given Borel set $U$ in $\R^n$ with $n \in \N$. The Lebesgue sets in 
$U$ are accordingly denoted by $\mathcal{B}^*(U)$. Also note that 
$0 \notin \N$. 

\section{Review of three notable theorems}
\label{sec:review}

This section lays the mathematical 
foundation for the analysis in later sections and also introduces the general 
physical context in which these results gain their relevance. 
In particular, we review two notable Bell inequalities as well as the closely related 
GH(S)Z Theorem. They may be considered ``representative'' of other 
theorems along the lines of Bell's original result \cite{bellEinsteinPodolskyRosen1964,
bellSpeakableUnspeakableQuantum2004}, 
in the sense that the line of reasoning employed there is usually akin to the ones 
used in the proofs of these three theorems.  

We first consider a slightly generalized version of Bell's original theorem \cite{bellEinsteinPodolskyRosen1964,bellSpeakableUnspeakableQuantum2004}. 
	\begin{Theorem}[Bell \cite{bellEinsteinPodolskyRosen1964}]
		\label{Thm:Bell}
		Let $\left(\Lambda, \mathcal{A}, \mathbb{P}\right)$ 
		be a probability space and let $S$ be any 
		nonempty set. Let 
				\begin{subequations}
				\label{eq:defAB}
				\begin{gather}
						A \colon S \times 
						\Lambda \to \lbrace -1, +1\rbrace \ , \quad 
						({a},\lambda) 
						\mapsto A(a)(\lambda) 
						\label{eq:defA}
						\\
						B \colon S \times 
						\Lambda \to \lbrace -1, +1\rbrace \ , \quad 
						({b},\lambda) 
						\mapsto B(b)(\lambda)
						\label{eq:defB}
				\end{gather}
				\end{subequations}
		be families of random variables such that 
		for all ${a} \in S$ the equality 
		\begin{equation}
			A(a) B(a) = -1
			\label{eq:Bellcorr}
		\end{equation}
		holds almost surely. 
		For any ${a},{b} \in S$, set 
		\begin{equation}
			\Evalue({a},{b})= \mathbb{E}\bigl(A(a) B(b)\bigr) \ .
			\label{eq:originalE} 
		\end{equation}
				
		Then for all ${a}, {b}, {c} \in S$ the following holds:  
		\begin{equation}
			\label{eq:Bellineq}
			1 + \Evalue({a},{b})  \bigr) 
				\geq \abs{\Evalue({a},{c}) 
					- \Evalue({c},{b})  \bigr)} \ .
		\end{equation}
	\end{Theorem}
	\begin{Proof}
		\begin{subequations}
		Due to boundedness, arbitrary products of $A$ and $B$ are integrable. 
		Moreover, Eq. \eqref{eq:Bellcorr} implies that 
		$B(a) = - A(a)$ almost surely. We may 
		therefore write 
		\begin{align}
			\Evalue\bigl({a}, {c} \bigr) 
					- \Evalue\bigl({c}, {b} \bigr)
						&= - \int_\Lambda \d \mathbb{P} \, 
							\left(A({a}) A({c})
								- A({c}) A({b}) \right)
								\label{eq:Bell_proofline}
						\\ 
						&=  - \int_\Lambda \d \mathbb{P} \, 
								A({a}) A({c})
							\left(1 
								- A({a}) A({b}) \right) \ .
		\end{align}
		Taking absolute values, applying the triangle inequality for integrals, 
		and recalling that $1 - A({a}) A({b}) \geq 0$, 
		the assertion follows. 
		\end{subequations}
	\end{Proof}
	
	Eq. \eqref{eq:Bellineq} is commonly called 
	\emph{Bell's inequality}. 

	In heuristic terms and for the 
	purpose of this article, we use the term \emph{Bell inequality} 
	to refer to any inequality in terms of $\Evalue(a,b)$, which 
	-- by means of an analogy -- can be violated by quantum 
	mechanical predictions. That this violation relies on an analogy 
	is due to the fact that quantum mechanics 
	relies on a different notion of probability and thus a different 
	notion of expectation value 
	(cf. Sec. 2 in \cite{reddigerApplicabilityKolmogorovsTheory2025}). 
	The most prolific example for the violation of a Bell inequality 
	in this sense is provided by the quantum-mechanical expectation value for a 
	$2$-body spin-$1/2$ singlet state 
	\parencites{bohmDiscussionExperimentalProof1957}
	{bellEinsteinPodolskyRosen1964}
	{bellSpeakableUnspeakableQuantum2004}
	{clauserProposedExperimentTest1969}
	{bellIntroductionHiddenvariableQuestion1971}\relax. 		
	
\begin{figure}[ht]
	\centering
	\includegraphics[width=0.8 \linewidth]{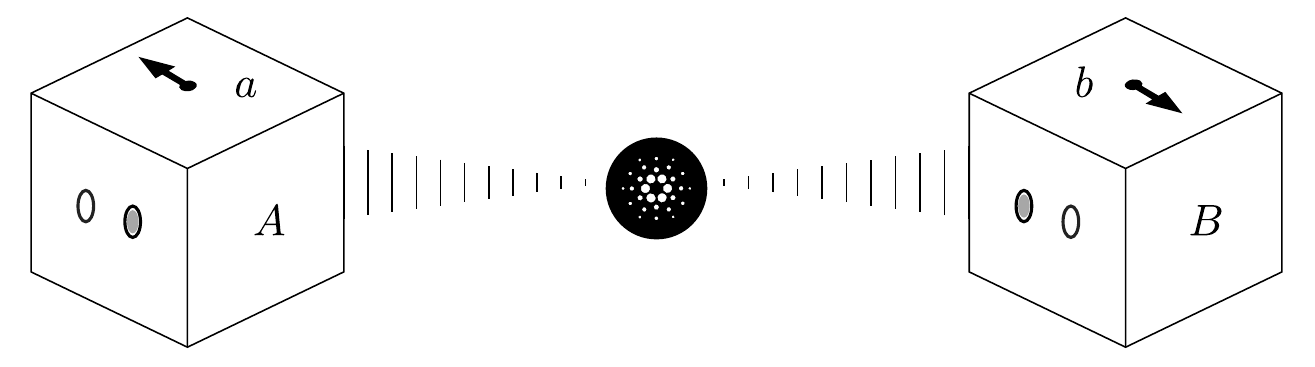}
	\caption{Sketch of the measurement setup for a general Bell experiment. In this 
				particular trial, 
				detector $A$ with setting $a$ records the value 
				$+1$. 
				Detector $B$ with setting $b$ records the value 
				$-1$.  Bell inequalities are tested by recording multiple outcomes 
				for various settings. 
				}
	\label{fig:Bellexperiment}
	\end{figure}

	We refer to any random experiment, which may be used to test 
	a Bell inequality, as a \emph{Bell experiment}. The usual setup for such an 
	experiment 
	is depicted in Fig. \ref{fig:Bellexperiment}: 
	Two similar detectors, also denoted by $A$ and $B$, 
	are placed apart from each other with a particle source in between. 
	In an idealized version of the experiment and in a brief amount of time 
	after triggering the 
	source, each detector records an output of either 
	$-1$ or $+1$. The probability of obtaining either value is different 
	for different detector settings. These are represented by parameters  
	$a, b \in S$. The outcome of the random experiment 
	is then described by the random variables $A(a)$ for detector $A$ 
	and $B(b)$ for detector $B$, respectively. That is, in a given trial with  
	given $\lambda$, 
	detector $A$ with setting $a$ records the value 
	$A(a)(\lambda)$ and detector $B$ with setting $b$ records the value 
	$B(b)(\lambda)$. It is noteworthy that Bell experiments may in 
	principle be carried out for a large variety of detectors 
	and for various particle sources, since no further 
	restrictions are placed on the random variables or the detector settings. 
	Furthermore, Bell experiments with more than two detectors and even 
	more than one source \cite{mjelvaDelayedchoiceEntanglementSwapping2024} 
	may be carried out, 
	but in this work we shall always assume the case of two detectors and a single 
	source, 
	unless noted otherwise. 
	
	In 1969 Bell's theorem was generalized by 
	Clauser et al. (cf. Eq. (1a) in \cite{clauserProposedExperimentTest1969}). 
	Their Bell inequality does not 
	require the \emph{correlation assumption} \eqref{eq:Bellcorr} and thus 
	overcomes the ``unrealistic restriction that for some pair of parameters [...] 
		there is perfect correlation'' \cite{clauserProposedExperimentTest1969}. 
	A closely related, more common inequality   
	was first shown by Bell in 
	1971 \cite{bellIntroductionHiddenvariableQuestion1971,
	bellSpeakableUnspeakableQuantum2004,
	clauserExperimentalConsequencesObjective1974}. The following Bell inequality is a corollary 
	of that result and is considered here, since   
	it eases the statistical analysis in Sec. 
	\ref{sec:statistics}. 
	\begin{Theorem}[Bell-Clauser-Horne-Shimony-Holt \cite{clauserProposedExperimentTest1969,
	bellIntroductionHiddenvariableQuestion1971,
	bellSpeakableUnspeakableQuantum2004}%
	]
		\label{Thm:BCHSH}
		Under the hypotheses of Thm. 
		\ref{Thm:Bell} yet dropping assumption 
		\eqref{eq:Bellcorr},   
		the following inequality is satisfied 
		for all ${a}, {b}, {a}', {b}'  \in S$: 
		\begin{equation}
			\abs{\Evalue({a},{b}) 
					+ \Evalue({a}',{b}') 
					+ \Evalue({a}',{b}) 
					- \Evalue({a},{b}') }
			\leq 2 \ .
			\label{eq:BCHSH}
		\end{equation}
	\end{Theorem}
	\begin{Proof}
		\begin{subequations}
		The proof follows Sec. 5.3.1 in the book by Moretti
	\cite{morettiFundamentalMathematicalStructures2019}. 
		See also \cite{gillStatisticsCausalityBells2014} for a slightly different proof. 
					
		We consider the left hand side of Eq. \eqref{eq:BCHSH}, use linearity of the integral 
		and apply the triangle inequality for integrals. Now, the following holds: 
		\begin{align}
			& \abs{A(a) \,  \bigl( B({b}) -  B(b')\bigr) 
				+ A(a') \, \bigl( B({b})
				+ B(b') \bigr) 
				} 
				\label{eq:BCHSH_proofline1}
			\\
			& \leq 
			\abs{A(a) \, \bigl( B({b}) -  B(b')\bigr) }
			+ \abs{ 
				A(a') \, \bigl( B({b})
				+ B(b') \bigr) 
				}
			\\ 
			 & \leq 
			\abs{B({b}) -  B(b') }
			+ \abs{ B({b})
				+ B(b') 
				}
				\label{eq:BCHSH_proofline2}
		\end{align}
		Since for $x,y \in \R$ with $\abs{x}, \abs{y} \leq 1$ 
		the term $\abs{x-y}+\abs{x+y}$ is   
		less or equal to $2$, 
		this also holds for the above expression.  
		\end{subequations}
	\end{Proof}
	Following \cite{morettiFundamentalMathematicalStructures2019}, we 
	henceforth refer to Eq. \eqref{eq:BCHSH} as the \emph{BCHSH inequality}. 
	It is more suitable for empirical 
	investigations than Bell's inequality, 
	since Eq. \eqref{eq:Bellcorr} does not need to hold. 
	
	The violation of the BCHSH inequality \eqref{eq:BCHSH}
	for the aforementioned $2$-body spin-$1/2$ singlet state
	is expressed by the following corollary. 
	\begin{Corollary}[\parencites{bellEinsteinPodolskyRosen1964}{clauserProposedExperimentTest1969}{bellIntroductionHiddenvariableQuestion1971}
	{bellSpeakableUnspeakableQuantum2004}]
		\label{Cor:BCHSHcontra}
		Let 
		$\left(\Lambda, \mathcal{A}, \mathbb{P}\right)$ 
		be a probability space and let the set  
		$S$ be the unit $2$-sphere $\mathbb{S}^2$ in $\R^3$. 
		Then there do not exist 
		families of random variables $A$, $B$, 
		as given by Eqs. \eqref{eq:defAB} above, 
		such that for all $\vec{a}$, $\vec{b} \in \mathbb{S}^2$ it holds that 
		\begin{equation}
			\Evalue(\vec{a},\vec{b}) = - \vec{a} \cdot \vec{b} \ .
			\label{eq:naiveQM}
		\end{equation}
	\end{Corollary}
	\begin{Proof}
		Assuming the contrary, 
		take any $\vec{a} \in \mathbb{S}^2$ and rotate it step-wise by an 
		angle of $\pi/4$ in a plane to get $\vec{b}$, $\vec{a}'$, and $\vec{b}'$, 
		respectively. Then Eq. \eqref{eq:BCHSH} is contradicted, since   
		\begin{equation}
			\abs{\Evalue(\vec{a},\vec{b}) 
					+ \Evalue(\vec{a}',\vec{b}') 
					+ \Evalue(\vec{a}',\vec{b}) 
					- \Evalue(\vec{a},\vec{b}')} = 2 \sqrt{2} \approx 2.828 \ .
		\end{equation}
	\end{Proof}
	
	\begin{Example}
		\label{Ex:Bell}
	In this ``classical'' model, we take 
	the ``spin vector'' of each particle to be a unit vector in 
	$\R^3$ 
	(cf. \cite{takabayasiVectorRepresentationSpinning1955,janossyHydrodynamicalModelWave1966}). 
	Here there 
	are simple, arguably natural choices for the (family of) random variables 
	$A$ and $B$ (cf. Eq. 4 in \cite{bellEinsteinPodolskyRosen1964} or 
	Chap. 2 of \cite{bellSpeakableUnspeakableQuantum2004}). 
	
	For the particle `measured' by detector $A$ the corresponding ``spin vector''
	is 
	\begin{equation}
						\vec{s}_A \colon  
						\Lambda \to \mathbb{S}^2 \ , \quad 
						\lambda  \mapsto 
						\vec{s}_A (\lambda)	\ . 
	\end{equation}
	Similarly, one considers $\lambda \mapsto \vec{s}_B (\lambda)$ for 
	detector $B$. 
	The aforementioned choices are then 
	\begin{subequations}
	\begin{equation}
		A(\vec{a}, \lambda ) =  
			\begin{cases} 
				\sgn\bigl( \vec{a} \cdot \vec{s}_A(\lambda) \bigr)	
					\phantom{-}	
					& , \ \vec{a} \cdot \vec{s}_A(\lambda) \neq 0\\
					1			
					& , \ \vec{a} \cdot \vec{s}_A(\lambda) = 0
			\end{cases}
			\label{eq:modelA}
	\end{equation}
	and 
	\begin{equation}
		B(\vec{b}, \lambda ) =  
			\begin{cases}
				\sgn\bigl( \vec{b} \cdot \vec{s}_B(\lambda) \bigr)	
					& , \ \vec{b} \cdot \vec{s}_B(\lambda) \neq 0\\
					-1			
					& , \ \vec{b} \cdot \vec{s}_B(\lambda) = 0
			\end{cases} 
			\ .
			\label{eq:modelB}
	\end{equation}
	If $s_B(\lambda) = - s_A (\lambda)$ for almost all $\lambda \in \Lambda$, 
	then the correlation assumption \eqref{eq:Bellcorr}, is 
	automatically satisfied. In that case Bell' inequality \eqref{eq:Bellineq} 
	holds. The BCHSH inequality \eqref{eq:BCHSH} holds independent of 
	the validity of Eq. \eqref{eq:Bellcorr}.  
	\end{subequations}
	\end{Example}

	The last theorem we consider in this section is known as (a variant of) 
	the GH(S)Z Theorem \cite{greenbergerGoingBellsTheorem1989,
	greenbergerBellsTheoremInequalities1990,merminWhatWrongThese1990,
	merminHiddenVariablesTwo1993,vaidmanVariationsThemeGreenbergerHorneZeilinger1999,
	brunnerBellNonlocality2014,brunnerPublishersNoteBell2014}. 
	It is meant to apply 
	to Bell experiments with four detectors 
	(cf. Fig. 2 in \cite{greenbergerBellsTheoremInequalities1990}). 
	The statement is similar to Cor. \ref{Cor:BCHSHcontra} in the sense that 
	it states that the quantum-mechanical expectation values for a 
	certain $4$-body spin-$1/2$ system 
	cannot be reproduced within the class of probabilistic models 
	specified below. A derivation of the respective  
	expectation values in quantum mechanics is given 
	in Appx. F of \cite{greenbergerBellsTheoremInequalities1990}. 
		
	\begin{Proposition}[Greenberger-Horne-Shimony-Zeilinger 
		\cite{greenbergerBellsTheoremInequalities1990}]
		\label{Prop:GHSZ}
		Let $\left(\Lambda, \mathcal{A}, \mathbb{P}\right)$ 
		be a probability space and denote by 
		$\mathbb{S}^1$ the quotient set $\R / 2 \pi \Z$ (i.e. the 
				set of angles up to the addition of integer multiples of $2 \pi$). 
				 
		Then there do not exist $1$-parameter 
		families of random variables $A$, $B$, $C$, $D$,  
		indexed by 
		$\alpha, \beta, \gamma, \delta \in \mathbb{S}^1$, respectively, and taking 
		values in $\lbrace -1, +1 \rbrace$, such that 
		\begin{equation}
			\mathbb{E}\bigl( A(\alpha) B(\beta) 
				C(\gamma) D(\delta) \bigr) 
				= - \cos \left(\alpha + \beta - \gamma - \delta \right) \ .
				\label{eq:GHSZ_exp}
		\end{equation}		
	\end{Proposition}
	\begin{Proof}
		This proof by contradiction 
		follows p. 1134 sq. in \cite{greenbergerBellsTheoremInequalities1990}. First, 
		observe that  
		\begin{multline}
			\mathbb{E}\bigl( A(\alpha) B(\beta) 
				C(\gamma) D(\delta) \bigr) 
			= 
			\mathbb{P} 
			\left( \set{\lambda \in \Lambda}{  A(\alpha, \lambda) B(\beta, \lambda) 
				C(\gamma, \lambda) D(\delta, \lambda) =1} \right) 
				\\
				- 
			\mathbb{P} 
			\left( \set{\lambda \in \Lambda}{  A(\alpha, \lambda) B(\beta, \lambda) 
				C(\gamma, \lambda) D(\delta, \lambda) =-1} \right)
				\ . 
				\label{eq:GHSZ_probsplit}
		\end{multline}	
	Therefore, whenever $\alpha + \beta - \gamma - \delta = 0$, Eq. 
	\eqref{eq:GHSZ_exp} implies that 
	\begin{equation}
		A(\alpha) B(\beta) 
				C(\gamma) D(\delta)  =-1
				\label{eq:GHSZ_-1}
	\end{equation}
	almost surely (the latter restriction was omitted in 
	\cite{greenbergerBellsTheoremInequalities1990}). 
	Hence, for all $\phi \in \mathbb{S}^1$ the following equations hold almost surely: 
	\begin{subequations}
	\begin{align}
		A(0) B(0) C(0) D(0) & =-1
		\label{eq:GHSZ_proof1}
		\\
		A(\phi ) B(0) C(\phi ) D(0) & =-1
		\label{eq:GHSZ_proof2}
		\\
		A(\phi ) B(0) C(0 ) D(\phi) & =-1
		\label{eq:GHSZ_proof3}
		\\
		A(2 \phi ) B(0) C(\phi ) D(\phi) & =-1 
		\label{eq:GHSZ_proof4}
		\ .
	\end{align}
	\end{subequations}
	We now multiply Eq. \eqref{eq:GHSZ_proof1} with Eq. \eqref{eq:GHSZ_proof2} 
	and Eq. \eqref{eq:GHSZ_proof1} with Eq. \eqref{eq:GHSZ_proof3}. The resulting 
	two equations we 
	again multiply with another, so that   
	\begin{equation}
		C(\phi ) D(\phi ) = C(0 ) D(0 )
	\end{equation}
	holds almost surely. Plugging this into Eq. \eqref{eq:GHSZ_proof4} and 
	then multiplying again with Eq. \eqref{eq:GHSZ_proof1}, we find that 
	$A(2\phi) = A (0)$ almost surely. Thus, $A(\pi) = A (0)$ almost surely.
	
	However, from Eqs. \eqref{eq:GHSZ_exp} and 
	\eqref{eq:GHSZ_probsplit} we also obtain that 
	\begin{equation}
		A(\pi) B(0) C(0) D(0) =1
	\end{equation}
	almost surely. Multiplying this again with Eq. 
	\eqref{eq:GHSZ_proof1}, we find that $A(\pi) = - A(0)$ almost surely, contrary 
	to our previous result. 
	\end{Proof}
	
Prop. \ref{Prop:GHSZ} is noteworthy as it directly shows 
the nonexistence of the respective random variables 
without relying on any Bell inequalities. 
Nonetheless, as the following remark clarifies, 
Prop. \ref{Prop:GHSZ} is only of minor relevance 
to the physical debate on ``hidden variables''. 
\begin{Remark}
\label{Rem:GHSZ}
	At first glance, Prop. \ref{Prop:GHSZ} may be viewed as the most direct 
	result of Sec. \ref{sec:review} on how probabilistic models 
	that are local in Bell's sense (Def. \ref{Def:locality} below) 
	conflict with 
	quantum-mechanical predictions. Yet, as acknowledged in the original references 
	\cite{greenbergerGoingBellsTheorem1989,
	greenbergerBellsTheoremInequalities1990}, the underlying argument
	crucially depends on the ``perfect correlation'' \eqref{eq:GHSZ_exp} 
	implied by said prediction. The 
	proof of Prop. \ref{Prop:GHSZ} reveals 
	that as soon as Eq. \eqref{eq:GHSZ_-1} only 
	holds with 
	very high probability -- not almost surely -- the argument fails. 
	Since Eq. \eqref{eq:GHSZ_exp} is an approximation to the empirical 
	behavior at best, the assumption of ``perfect correlation'' constitutes an 
	overidealization \cite{gillStatisticsCausalityBells2014}. 
	Indeed, as Gill  \cite{gillStatisticsCausalityBells2014} 
	comments sharply, ``the logical conclusion 
	from the experiment is that nothing has been proved'' and that, 
	for this reason ``whatever experimental set-up we take'' 
	Bell inequalities will need to be considered instead. 
	
	In this respect, it is also worth pointing out that 
	Prop. \ref{Prop:GHSZ} does not account for 
	``contextuality''` \cite{khrennikovContextualistViewpointGreenberger2001}, 
	as discussed in Sec. \ref{sec:generalizedBell} 
	for Bell experiments with two detectors. 
\end{Remark}

\section{Bell models and locality}
\label{sec:discussion}

This section provides a more in-depth physical analysis of the 
BCHSH inequality \eqref{eq:BCHSH} using methods and concepts from 
the mathematical theory of probability. In doing so, we also  
elaborate on the concept of locality in Bell experiments, which are described by 
so-called ``noncontextual'' Bell models. 
We primarily consider the BCHSH inequality, since it 
is stronger than Bell's inequality and does not suffer from the 
drawbacks of the GH(S)Z-Theorem, as discussed in Rem. \ref{Rem:GHSZ}.
	
Here the primary mathematical object of study is 
what we call a Bell model. It constitutes a slight 
generalization of the assumptions of the BCHSH theorem, Thm. \ref{Thm:BCHSH}. 
\begin{Definition}
\label{Def:Bellmodel}
\begin{subequations}
	\label{eq:defAB_gen}
	Let $\left(\Lambda, \mathcal{A}, \mathbb{P}\right)$ 
	be a probability space, let $S$ be any 
	nonempty set and let 
	\begin{gather}
		A \colon S \times S \times 
		\Lambda \to \lbrace -1, +1\rbrace \ , \quad 
		(a,b,\lambda) 
		\mapsto A(a,b)(\lambda) 
		\label{eq:defAgen}
		\\
		B \colon S \times S \times 
		\Lambda \to \lbrace -1, +1\rbrace \ , \quad 
		(a,b,\lambda) 
		\mapsto B(a,b)(\lambda)
		\label{eq:defBgen}
	\end{gather}
	be two families of random variables. 
	We call the tuple 
	$\left(\Lambda, \mathcal{A}, \mathbb{P}, S, A, B\right)$ a \emph{Bell model}. 
\end{subequations}
\end{Definition}

The primary purpose of the above concept is to study Bell experiments: 
for a given $\lambda$ and setting parameters 
$a$ and $b$, detector $A$ records the number $A(a,b)(\lambda)$ and detector $B$ the 
number $B(a,b)(\lambda)$, in full analogy to the situation in 
Fig. \ref{fig:Bellexperiment}. 

The setting parameters $a$ and $b$ are assumed to be chosen freely 
by the experimenter. Since we would like to discuss the respective 
BCHSH inequality \eqref{eq:BCHSH} and that inequality becomes trivial for either 
$a =a'$ or $b=b'$, $S$ should contain at least $2$ distinct parameters 
(e.g. $a=b \neq a' =b'$ may yield $4$ distinct summands). 
It is also possible to add an additional layer of randomness by choosing 
the detector settings randomly; indeed, some authors start with this case and 
then consider probability measures conditioned on certain settings---which 
often amounts to an overcomplication. 
The case of random detector settings will be covered implicitly 
in Sec. \ref{sec:generalizedBell} below and shall not be of concern here.  

In a given trial, it is always the 
product of the two readings 
\begin{equation}
	X({a}, {b})(\lambda) = A({a}, {b})(\lambda) \, 
			B({a}, {b})( \lambda)  \ . ´
			\label{eq:defX_general}
\end{equation}
which is recorded. That we consider the product of two random variables in each trial 
is not an assumption, but an experimental choice. As Lambare notes 
(cf. \cite{lambareBellInequalitiesCounterfactual2021} and 
Sec. 4.1 in \cite{lambareQuestioningReasonablenessQuantum2026}), some 
treatments in the literature do not take account of this (see e.g. 
\cite{fineHiddenVariablesJoint1982,rastallBellInequalities1983} 
as well as Appx. C in \cite{nagasawaLocalityHiddenvariableTheories1997} commenting 
on \cite{delapenaHiddenvariableTheoriesBell1972}). 

This raises the question of the physical 
interpretation of $\lambda$. 

We shall first recall some basic concepts 
from the mathematical theory 
of probability. The set $\Lambda$ is known as the \emph{sample space}. Accordingly, any element 
$\lambda$ is called a \emph{sample} or \emph{elementary event}. By assumption, in a particular trial of the random experiment of interest, the 
sample $\lambda$ is chosen randomly and realized in the sense that 
a particular value is picked. To compute the probabilities of obtaining 
particular values, we may choose a \emph{random event} $U$ from the \emph{collection 
of random events} $\mathcal{A}$. By construction of $\mathcal{A}$, $U$ is a subset of 
$\Lambda$. $\mathbb{P}(U)$ is then the probability that in a particular trial 
any sample $\lambda$ contained in $U$ is realized. 

In a Bell experiment, $\lambda$ is only 
indirectly as well as impartially deducible through the detector readings. 
In other words, the experimenter only has access to the 
numbers $A(a,b)(\lambda)$ and $B(a,b)(\lambda)$ through the detectors, not 
the sample $\lambda$ itself. It is in this sense that $\lambda$ is indeed 
a ``hidden variable''. 

Of course and as stated in Sec. \ref{sec:intro}, in contemporary quantum theory 
the term ``hidden variable'' carries a slightly different connotation: 
they aim to provide a more detailed description of quantum systems. 

The study of Bell inequalities in this context is justified in 
the sense that, at least in principle, much physics can be
incorporated into $\lambda$: 
it may include initial conditions 
for point particles, but possibly also information on various different field configurations 
\cite{bellEinsteinPodolskyRosen1964,bellIntroductionHiddenvariableQuestion1971,
bellSpeakableUnspeakableQuantum2004}. 
If we indeed follow this route, then the set $\Lambda$ ought to be uncountably infinite. 
Furthermore, the probability $\mathbb{P}(\lbrace \lambda \rbrace)$, 
that a particular $\lambda \in \Lambda$ is realized, 
ought to be zero for any choice of $\lambda$ (assuming  
$\lbrace \lambda \rbrace \in \mathcal{A}$). The reader may find motivating examples  
for instance in \cite{dewdneySpinNonlocalityQuantum1988,beyerSternGerlachEPRB2024,
lombardiniInteractingQuantumTrajectories2024,arroyoFamilyDeterministicModels2025}. 
The paradox that any one of the events of probability zero is sure to occur 
should not raise any concerns, for it is common, for instance, in the 
mathematical study of Brownian motion (cf. Rem. 2 in \S 2 of \cite{kolmogorovFoundationsTheoryProbability1956}). Indeed, 
in the theory of stochastic mechanics Wiener processes are used to model 
particle trajectories \cite{nelsonDerivationSchrodingerEquation1966,
nelsonQuantumFluctuations1985,loffredoLagrangianVariationalPrinciple1989,
nagasawaStochasticProcessesQuantum2000,
farisIntroductionDiffusiveMotion2006,
carlenStochasticMechanicsLook2006,delapenaEmergingQuantumPhysics2015,
beyerStochasticMechanicsFoundation2021,kuipersQuantumMechanicsStochastic2023,
beyerSternGerlachEPRB2024}.

Still, from the point of view of mathematical probability theory, usage of the term 
``hidden variable'' for the sample $\lambda$ in a Bell experiment may be considered 
problematic. For instance, in the usual model for the throw of a fair $6$-sided die 
no mechanical explanation for the probabilities is and needs to be given---it 
suffices to name each sample $1, \dots, 6$ and assign a corresponding probability. Indeed, geometric quantum theory takes a similar approach and 
in this sense it is not a ``hidden variable theory'' 
(see e.g. \cite{reddigerMadelungPictureFoundation2017,
				reddigerProbabilisticFoundationNonrelativistic2022,
				reddigerApplicabilityKolmogorovsTheory2025,
				reddigerAddendumApplicabilityKolmogorovs2026}). 
			
Having clarified the basic ingredients of a Bell model, we proceed with our analysis. 

First, the following implicit assumption is to be pointed out: we only 
consider Bell experiments, where both detectors register a signal. 
Of course, one may in principle choose a convention, where the value $-1$ or $+1$ is 
assigned, even if the respective detector does not provide any reading. Yet, since this is a rather contrived practice, Bell models in the sense of Def. \ref{Def:Bellmodel} ought not to 
be used for the description of a Bell experiment, if nondetection does occur. 
The following remark clarifies the (non)validity of the BCHSH inequality in that case. 
	\begin{Remark}	
		\label{Rem:BCHSHothervalues}
		The proof of Thm. \ref{Thm:BCHSH} shows 
		that the BCHSH inequality \eqref{eq:BCHSH} also holds, if 
		we allow $A$ and $B$ to take any value in the closed interval $[-1,1]$ 
		(cf. Thm. 1 on p. 213 in 
		\cite{khrennikovProbabilityRandomnessQuantum2016}). 
		
		In particular, we may consider the case that they take 
		values in the set $\lbrace -1, 0 , +1 \rbrace$ and assign the number  
		$0$ to the respective random variable, if that detector registers no signal 
		\cite{jarrettPhysicalSignificanceLocality1984,
		gargDetectorInefficienciesEinsteinPodolskyRosen1987,
		larssonBellsInequalityDetector1998,avisSingleCompleteProbability2009,
		khrennikovCHSHInequalityQuantum2015}. 
		Since we are only interested in those trials, where 
		both detectors register a signal, we need to formally 
		condition on the event
		\begin{subequations} 
		\begin{align}
			\Lambda_{ab} 
			& = \lbrace A(a) \neq 0 \ \text{and} \ B(b) \neq 0 \rbrace 
			\\
			&= \set{\lambda \in \Lambda \,}{\, A(a)(\lambda) \neq 0 \ 
			\text{and} \ B(b)(\lambda) \neq 0}
			\ .
		\end{align}
		\end{subequations} 
		The set $\Lambda_{ab}$ 
		generally depends on both setting parameters, $a$ and $b$. Hence, the respective 
		conditional probability measure also depends thereon. Under the 
		assumptions of Thm. \ref{Thm:BCHSH}, the 
		respective conditional expectation value is given by 
		\begin{equation}
			\Evalue({a},{b}) 
			= \frac{1}{\mathbb{P}(\Lambda_{ab})} 
				\int_{\Lambda_{ab}} A(a) B(b) \, \d \mathbb{P} \ , 
		\end{equation}
		assuming $\mathbb{P}(\Lambda_{ab}) \neq 0$.
		For this choice of $\Evalue({a},{b})$ the BCHSH inequality need not hold, 
		though in Sec. V of \cite{larssonBellsInequalityDetector1998} 
		an adapted inequality has been derived. In the literature, statistical 
		violations of the BCHSH inequality due to 
		this phenomenon are known as the ``detection loophole''
		\cite{brunnerBellNonlocality2014,brunnerPublishersNoteBell2014}. 
		The interested reader is also referred to  
		\cite{larssonBellsInequalityCoincidencetime2004,
		avisSingleCompleteProbability2009,
		gillStatisticsCausalityBells2014,
		khrennikovCHSHInequalityQuantum2015}. 
		
		Indeed, the use of 
		a setting-dependent conditional probability measure may be 
		employed as a further motivation for 
		the introduction of contextual Bell models 
		in Sec. \ref{sec:generalizedBell} below. 
		Using the usual 
		notation for the conditional probability measure, we may write 
		\begin{equation}
			\Evalue({a},{b}) 
			= \int_{\Lambda} A(a) B(b) \, \d 
			\mathbb{P}\left( \, . \, \vert \Lambda_{ab}\right)  \ .			
		\end{equation}
	\end{Remark}

In the mathematical study of Bell models, there are two natural notions of 
``locality''. Both notions can be traced back to Bell's 
original works \cite{bellEinsteinPodolskyRosen1964,bellTheoryLocalBeables1976,
bellSpeakableUnspeakableQuantum2004}, 
though respective definitions in the literature do often not meet mathematical 
standards of rigor. Since one of these notions is weaker than the other, we shall 
choose the 
respective terms ``weak'' and ``strong locality''.  
	\begin{Definition}
		\label{Def:locality}
		Let $\left(\Lambda, \mathcal{A}, \mathbb{P}, S, A, B\right)$ 
		be a Bell model. 
		\begin{enumerate}[1)]
			\item 
			\begin{subequations}
				The Bell model is called \emph{strongly local} (or  
					\emph{Bell local}), if 
					for all $a, a', b, b' \in S$
					\begin{equation}
						A(a,b)= A(a,b') 
					\end{equation}	
						almost surely and 
					\begin{equation}
						B(a,b)= B(a',b)  
					\end{equation}
					almost surely. 
			\end{subequations}
			\item 
			\begin{subequations}
				The Bell model is called \emph{weakly local}, if 
					for all $a, a', b, b' \in S$ 
					the respective probability distributions satisfy 
					\begin{equation}
						\mathbb{P} \circ \bigl(A(a,b) \bigr)^{-1} = 
							\mathbb{P} \circ \bigl(A(a,b') \bigr)^{-1} \ \phantom{.}
					\end{equation}						
					and 
					\begin{equation}
						\mathbb{P} \circ \bigl(B(a,b) \bigr)^{-1} = 
							\mathbb{P} \circ \bigl(B(a',b) \bigr)^{-1} \ . 
					\end{equation}	
			\end{subequations}					
		\end{enumerate}
	\end{Definition}
	It directly follows from the above definition that 
	a Bell model is strongly local, if and only if there exist families of random variables 
	$\tilde{A}$ and $\tilde{B}$ such that
	\begin{equation}
		A(a,b) = \tilde{A}(a) \quad \text{and} \quad B(a,b) = \tilde{B}(b)
		\label{eq:reducetoBellwhenstrong}
	\end{equation}
	almost surely for all $a,b \in S$. 
	
	Similarly, a Bell model is weakly local, if and only if the equations 
	\begin{subequations}
		\label{eq:defpApB}
	\begin{align}
		p_{A\pm}(a) &= \mathbb{P}\bigl( A({a},{b}) = \pm 1 \bigr) 
		\\
		p_{B\pm}(b) &= \mathbb{P}\bigl(B({a},{b})= \pm 1\bigr) 
	\end{align}
	\end{subequations}
	determine well-defined families of probabilities 
	$\bigl(p_{A\pm}(a)\bigr)_{a \in S}$ and $\bigl(p_{B\pm}(b)\bigr)_{b \in S}$. 
	
	Building upon Def. \ref{Def:locality}, we may rigorously define  
	\emph{strong locality} (or \emph{Bell locality}) as the property of a 
	Bell model to be strongly local (or Bell local). 
	Accordingly,  we define \emph{weak locality} as the property of a Bell model to be 
	weakly local. 
	
	In the context of Bell experiments, 
	these two notions (as well as their generalizations in 
	Def. \ref{Def:generalizedBM_locality} 
	below) enable the scientific study of 
	the otherwise metaphysical concept of ``locality''. 
	It remains to provide some 
	physical justification for this choice of terminology. 
	
	Strong locality is directly motivated from Bell's theorem \ref{Thm:Bell} 
	and the BCHSH theorem \ref{Thm:BCHSH}. For this purpose, we redefine  
	\begin{equation}
		\Evalue({a},{b})= \mathbb{E}\bigl(X({a}, b)\bigr) = 
		\mathbb{E}\bigl(A({a}, b) \, B({a},b)\bigr) \ . 
		\label{eq:newEab}
	\end{equation}
	Strong locality implies that we may effectively 
	drop the dependence on the other detector setting (cf. 
	Eq. \eqref{eq:reducetoBellwhenstrong}), so we may indeed 
	identify $\Evalue({a},{b})$ from Eq. \eqref{eq:newEab} 
	with the one from \eqref{eq:originalE}. Therefore, 
	the BCHSH theorem \ref{Thm:BCHSH} directly implies and 
	is indeed equivalent to the following statement. 
	\begin{Corollary}
		\label{Cor:stronglocalityBCHSH}
		Let $\left(\Lambda, \mathcal{A}, \mathbb{P}, S, A, B\right)$ 
		be a strongly local Bell model and define $\Evalue$ via Eq. 
		\eqref{eq:newEab} above. 
	
		Then the BCHSH inequality \eqref{eq:BCHSH} is satisfied. 
	\end{Corollary}
	
	Colloquially, weak locality corresponds to the following property: 
	if we consider a single detector only, the probabilities for that detector 
	only depend on its setting, not on the setting of the other detector 
	\cite{ballentineBellsTheoremDoes1987}. 
	In the literature, weak locality is often called 
	``no signaling'' 
	(cf. Sec. 7 in \cite{bellTheoryLocalBeables1976}, Chap. 7.7 in 
	\cite{bellSpeakableUnspeakableQuantum2004}): a change 
	in the setting of one detector can not be statistically inferred from the 
	readings of the other detector. In this sense, weak locality may be viewed as a  
	more tangible notion of locality than strong locality. 

	Though it is immediate from Def. \ref{Def:locality} that strong locality implies 
	weak locality, weak locality does not imply strong locality. The 
	following example illustrates this point. 
\begin{Example}
	\label{Ex:weakvsBell}
	Consider the probability space 
	$\left(\mathbb{S}^2, \mathcal{B}^*(\mathbb{S}^2), \mathbb{P}\right)$ 
	with $\mathbb{P}(U)=\operatorname{area}(U)/\operatorname{area}(\mathbb{S}^2)$. 
	That is, we take the uniform distribution on the $2$-sphere 
	$\mathbb{S}^2$ 
	(cf. Ex. 1.75 in \cite{klenkeProbabilityTheoryComprehensive2020}). 
	Denote the standard group action 
	of the 
	special orthogonal group $\operatorname{SO}_3$ on $\mathbb{S}^2$ via left 
	multiplication as follows: 
	\begin{equation}
		L_{\,.\,} \colon \operatorname{SO}_3 \times \ \mathbb{S}^2 \to \mathbb{S}^2 \ , 
		\quad (\mathbf{a},\vec{\lambda}) \mapsto L_\mathbf{a}(\vec{\lambda})= \mathbf{a} \cdot\vec{\lambda} \ . 
	\end{equation}
	Then for any $\mathbf{a} \in \operatorname{SO}_3$ we have 
	$\mathbb{P}\circ L_\mathbf{a}^{-1}=\mathbb{P}$. That is, 
	the probability measure is invariant under the group action. 
		
	Taking $S=\operatorname{SO}_3$, we shall construct a pair of not strongly 
	local, yet weakly local 
	random variables $A$ and $B$ in the sense of Eq. 
	\eqref{eq:defAB_gen} above. 
	
	Denoting the first column of a matrix $\mathbf{c} \in \operatorname{SO}_3$ 
	as $\vec{c}_1$, 
	for $\mathbf{a}, \mathbf{b} \in \operatorname{SO}_3$ and 
	$\vec{\lambda} \in \mathbb{S}^2$ these random variables are defined as follows: 
	\begin{subequations} 
	\begin{align}
		A \bigl(\mathbf{a},\mathbf{b})(\vec{\lambda}\bigr) &= 
				\begin{cases} 
				\sgn\bigl( \vec{a}_1 \cdot (\mathbf{b} \cdot \vec{\lambda}) \bigr)	
					\phantom{-}	
					& , \  \vec{a}_1 \cdot (\mathbf{b} \cdot \vec{\lambda}) \neq 0\\
					1			
					& ,  \ \vec{a}_1 \cdot (\mathbf{b} \cdot \vec{\lambda}) = 0
			\end{cases}
			\ ,
			\\
		B \bigl(\mathbf{a},\mathbf{b})(\vec{\lambda}\bigr) &= 
				\begin{cases} 
				- \sgn\bigl( \vec{b}_1 \cdot (\mathbf{a} \cdot \vec{\lambda}) \bigr)	
					& , \  \vec{b}_1 \cdot (\mathbf{a} \cdot \vec{\lambda}) \neq 0\\
					- 1			
					& ,  \ \vec{b}_1 \cdot (\mathbf{a} \cdot \vec{\lambda}) = 0
			\end{cases}
			\ .
	\end{align}
	\end{subequations} 
	
	The Bell model $\bigr(\mathbb{S}^2, \mathcal{B}^*(\mathbb{S}^2), \mathbb{P},
	\operatorname{SO}_3, A,B \bigl)$
	is not strongly local, 
	because the dependence on neither one of the settings is trivial. 
	
	However, using the construction in Ex. \ref{Ex:Bell}, 
	we obtain a family of random variables 
	$C \colon \operatorname{SO}_3 \times \ \mathbb{S}^2 \to 
	\lbrace -1, +1 \rbrace$ such that we may always write 
	\begin{equation}
			A \bigl(\mathbf{a},\mathbf{b})(\vec{\lambda}\bigr) = 
		C \bigl(\mathbf{a},L_\mathbf{b}(\vec{\lambda})\bigr)  
		\quad \text{and} \quad 
		B \bigl(\mathbf{a},\mathbf{b}, \vec{\lambda}\bigr) = - 
		C \bigl(\mathbf{b},L_\mathbf{a}(\vec{\lambda})\bigr)	\ . 
	\end{equation}
	Therefore, 
	\begin{align}
		p_{A+}(\mathbf{a}) 
			&= \mathbb{P} 
				\left( C \bigl(\mathbf{a},L_\mathbf{b}( \, . \, ) \bigr)^{-1}  
				\lbrace +1 \rbrace \right) \\
			&= \left( \mathbb{P} \circ L_\mathbf{b}^{-1} \right) 
				\left( C \bigl(\mathbf{a},\, . \,  \bigr)^{-1}  
				\lbrace +1 \rbrace \right) \\
			& =\mathbb{P} 
				\left( C \bigl(\mathbf{a},\, . \,  \bigr)^{-1}  
				\lbrace +1 \rbrace \right) 
	\end{align}
	is well-defined. An analogous argument applies to  
	$p_{B+}(\mathbf{b})$. Hence, the Bell model is weakly local. 
	Indeed, $p_{A+}(\mathbf{a}) \equiv p_{B+}(\mathbf{b}) \equiv 1/2$ 
	due to symmetry. 
\end{Example}

In the context of Bell experiments, we generally assume that weak 
locality holds. If there was substantial evidence in favor of a violation of weak locality 
in any Bell experiment of physical relevance, 
this would plausibly shift the physical debate on Bell inequalities into a 
qualitatively different direction. 

We may infer a quantitative relationship between 
strong locality and weak locality by considering the four  
joint probabilities: 
\begin{align}
	p_{\pm \mp}(a,b) 
		= \mathbb{P}\bigl( A(a,b) = \pm 1 \ \text{and} \ B(a,b) = \mp 1 \bigr) 
	\label{eq:defpjoint}
\end{align}
for all possible sign combinations. The following proposition describes 
this relationship.  
\begin{Proposition}
	\label{Prop:jointprobs}
	Let $\left(\Lambda, \mathcal{A}, \mathbb{P}, S, A, B\right)$ 
	be a weakly local Bell model. Define $p_{A\pm}$, $p_{B\pm}$, 
	$p_{\pm \mp}$ as well as $\Evalue$ via Eqs. \eqref{eq:defpApB}, 
	\eqref{eq:defpjoint} and \eqref{eq:newEab} above.  

	Then the following identities hold for all $a,b \in S$: 
	\begin{subequations}
	\label{eq:probidentities}
	\begin{align}
		p_{++}(a,b) &= -\frac{1}{4} + \frac{1}{2} \, p_{A+}(a) + 
			\frac{1}{2} \, p_{B+}(b) + \frac{1}{4} \, \Evalue(a,b) \\
		p_{+-}(a,b) &= \phantom{-}\frac{1}{4} + \frac{1}{2} \, p_{A+}(a) -\frac{1}{2} \, p_{B+}(b) -\frac{1}{4} \, \Evalue(a,b) \\
		p_{-+}(a,b) &= \phantom{-} \frac{1}{4} -\frac{1}{2} \, p_{A+}(a) + \frac{1}{2} \, p_{B+}(b) -\frac{1}{4} \, \Evalue(a,b) \\
		p_{--}(a,b) &= \phantom{-} \frac{3}{4} -\frac{1}{2} \, p_{A+}(a) -\frac{1}{2} \, p_{B+}(b) + \frac{1}{4} \, \Evalue(a,b) 
	\end{align}
	\end{subequations} 
\end{Proposition}
\begin{Proof}
	We obtain the following identities from elementary rules of probability 
	as well as an explicit consideration of the expectation value 
	$\Evalue(a,b)$: 
	\begin{subequations}
	\begin{gather}
		p_{A+}(a) = p_{++}(a,b) + p_{+-}(a,b) 
		\label{eq:proof_jointprobs_1}
		\\
		p_{B+}(b) = p_{++}(a,b) + p_{-+}(a,b) 
		\label{eq:proof_jointprobs_2}
		\\
		1 = p_{++}(a,b) + p_{+-}(a,b) + p_{-+}(a,b) + p_{--}(a,b) \\
		\Evalue(a,b) = p_{++}(a,b) - p_{+-}(a,b) - p_{-+}(a,b) + p_{--}(a,b)
		\label{eq:proof_jointprobs_4}
	\end{gather}
	\end{subequations}
	Eqs. \eqref{eq:probidentities} are then derived by solving the 
	respective system of linear equations. 
\end{Proof}
The probabilities $p_{\pm \mp}(a,b)$ 
describe a Bell model entirely in the 
following sense: on this level of description, 
two Bell experiments governed by the same probabilities for 
the different parameter settings are empirically 
indistinguishable. We call such Bell models 
\emph{isomorphic}. We shall not consider any rigorous definition here, for a 
rough concept suffices for the purpose of this work. 

Prop. \ref{Prop:jointprobs} shows that two weakly local Bell models are isomorphic, 
if the respective values for $p_{A+}(a)$, $p_{B+}(b)$, and $\Evalue(a,b)$ coincide in an 
appropriate sense for the different settings. These three quantities for 
different choices of settings may be viewed as 
the empirically relevant  ``independent variables'' of a Bell model. Strong locality 
then merely constitutes a constraint on $\Evalue(a,b)$, as provided, for instance, by   
the BCHSH inequality \eqref{eq:BCHSH} (cf. Cor. \ref{Cor:stronglocalityBCHSH}). 

We may thus summarize the relationship between weak locality and 
strong locality as follows. Strong locality implies weak locality and is hence, indeed, a 
stronger notion of locality. Conversely, weak locality does not imply strong locality, 
because the value of $\Evalue(a,b)$ cannot be deduced from the marginal 
probabilities $p_{A+}(a)$ and $p_{B+}(b)$ of either detector. Colloquially put, 
while weak locality is a property of each individual detector, 
strong locality is a property of the system as whole. 

Next we shall consider the correlation between the 
random variables $A(a,b)$ and $B(a,b)$ for given $a,b \in S$ in 
a weakly local Bell model. By definition, the two  
are \emph{correlated}, if the covariance 
	\begin{equation}
		\Cov \big(A(a,b),B(a,b)\bigr)  = 
		\Evalue(a,b) - \mathbb{E}\big(A(a,b)\bigr) \, \mathbb{E}\big(B(a,b)\bigr)
		\label{eq:covAB}
	\end{equation}
is nonzero. We also recall that they are stochastically independent, if 
\begin{align*}
	p_{\pm \mp} (a,b) = p_{A \pm}(a) \ p_{B \mp}(b) 
\end{align*}
and stochastically dependent otherwise. 
It is elementary that stochastically independent random variables 
are uncorrelated, but the reverse implication does generally not hold 
(see e.g. Sec. 5.1 in \cite{klenkeProbabilityTheoryComprehensive2020}).  

The following lemma shows that in a weakly local Bell model 
the BCHSH inequality \eqref{eq:BCHSH} can only be 
violated, if for some $a,b \in S$ the two random variables are indeed correlated (see also 
\cite{bellBertlmannsSocksNature1981,bellSpeakableUnspeakableQuantum2004,
hallSignificanceMeasurementIndependence2016}). 
	\begin{Lemma}
		\label{Lem:uncorrelated+weaklylocal}
		Let $\left(\Lambda, \mathcal{A}, \mathbb{P}, S, A, B\right)$ 
		be a weakly local Bell model and define $\Evalue$ as in    
		Eq. \eqref{eq:newEab} above. 
		
		If $A(a,b)$ and $B(a,b)$ are uncorrelated for all $a,b \in S$, then 
		the BCHSH inequality \eqref{eq:BCHSH} holds. 
	\end{Lemma}
	\begin{Proof}
		Since 
		\begin{align}
			\mathbb{E}\big(A(a,b)\bigr) = p_{A+}(a) - p_{A-}(a) = 2 p_{A+}(a) - 1\\ 
			\mathbb{E}\big(B(a,b)\bigr) = p_{B+}(b) - p_{B-}(b) = 2 p_{B+}(b) - 1
		\end{align}
		we may write $\Evalue_A(a)$ and $\Evalue_B(b)$ for the respective 
		expectation value. By assumption, the covariance in Eq. \eqref{eq:covAB} 
		vanishes. Hence, 
		\begin{equation}
			\Evalue(a,b) = \Evalue_A(a) \Evalue_B(b)\ . 
		\end{equation}
		We may therefore rewrite 
		the left hand side of the BCHSH inequality 
		\eqref{eq:BCHSH} as follows:  
		\begin{subequations}
		\begin{align}
			 & \abs{\Evalue_A(a) \Evalue_B(b) 
					+ \Evalue_A(a') \Evalue_B(b') 
					+ \Evalue_A(a') \Evalue_B(b)  
					- \Evalue_A(a) \Evalue_B(b')}
			\\
			& \leq  
			\abs{\Evalue_A(a)  \, \bigl( \Evalue_B(b) - \Evalue_B(b' )\bigr) }
			+ \abs{ 
				\Evalue_A(a') \, \bigl( \Evalue_B(b) + \Evalue_B(b' )\bigr)
				}
			\\ 
			& \leq 
			\abs{\Evalue_B(b) - \Evalue_B(b' )}
			+ \abs{ \Evalue_B(b) + \Evalue_B(b' )
				}	
			\\
			& \leq 2	
		\end{align}
	\end{subequations}
	\end{Proof}
	Conversely, 
	Lem. \ref{Lem:uncorrelated+weaklylocal} implies that 
	in a weakly local Bell model a violation 
	of the BCHSH inequality can only occur, if for some of the chosen settings 
	the detector random variables are correlated. In particular, they 
	must be stochastically dependent on each other. 
	
	With regards to the relationship between strong locality and stochastic 
	independence of the detector random variables two points 
	shall be raised. First,  
	strong locality does not imply stochastic independence of the 
	detector random variables, even though the literature has not always 
	been sufficiently clear on this point (cf. 
	Sec. 9 in \cite{bellNouvelleCuisine1990} or Chap. 24.9 in 
	\cite{bellSpeakableUnspeakableQuantum2004} as well as 
	\cite{ballentineBellsTheoremDoes1987,
	brownBellBellsTheorem2016,
	nagerStrongerBellArgument2020,
	lucWhatAreBearers2026}).  
	Second, if in a weakly local Bell model the detector random variables are 
	stochastically independent, then there exists a strongly 
	local Bell model to which it is isomorphic 
	(by defining new detector random 
	variables via 
	the sets $\lbrace A(a,b') = \pm 1 \rbrace$ and 
	$\lbrace B(a',b) = \pm 1 \rbrace$ for some fixed $a',b' \in S$). 

	The above analysis points to a close relationship between 
	the correlation of the detector random variables and the 
	quantum mechanical concept of ``entanglement''. Indeed, it is a 
	central question in the so-called Einstein-Podolsky-Rosen debate
	\cite{einsteinCanQuantumMechanicalDescription1935}, if and in what sense 
	``entanglement'' differs qualitatively from mere 
	correlation \cite{keylFundamentalsQuantumInformation2002,
	gisinCanRelativityBe2015,
	fryHistoryLeadingBells2019,
	janasUnderstandingQuantumRaffles2022,freirejuniorAlainAspectExperiments2022,
	sontzNewApproachQuantum2025}. 

	\section{Contextual Bell models and locality}	
	\label{sec:generalizedBell}
	
	In 
	this section we consider the question  
	whether the empirical violation of the BCHSH inequality may be explained  
	by including 
	``supplementary parameters'' 
	\cite{kupczynskiCanWeClose2017,kupczynskiEinsteinianNosignallingViolated2017} 
	for the detectors. 
	That is, does the BCHSH inequality still hold, if the detectors themselves 
	behave randomly and in accord with different probability distributions 
	for different detector settings? In the literature it was previously suggested that 
	the supposed failure of Bell inequalities to account for this possibility 
	opens up a so-called ``contextuality loophole'' in Bell experiments 
	\cite{nieuwenhuizenWhereBellWent2009,
	kupczynskiCanWeClose2017,kupczynskiEinsteinianNosignallingViolated2017}. 
	Here we shall elaborate on this statement by laying the basis for a more rigorous 
	study of contextuality in this setting. We will see that the notion of  
	contextuality introduced here does not merely constitute another 
	``loophole'', but that it 
	reveals a conceptual misunderstanding in terms of the 
	range of applicability of ``noncontextual'' Bell models, as studied 
	in the previous section. 
	For this purpose, we first motivate the mathematical definition of 
	a contextual Bell model and then carry over the  
	notions of strong and weak locality from Sec. \ref{sec:discussion}. 
	We will see how even 
	``classical'' Bell experiments motivate the introduction of 
	such models. Still, we will also see that 
	a wide class of contextual Bell models -- though not all -- does satisfy 
	the BCHSH inequality and work out a necessary 
	condition for this violation to occur. 
	
	We begin our motivation 
	with the requirement that the sample space $\Lambda$ in a given Bell experiment 
	may be written as a product of three sets:  
	\begin{equation}
				\Lambda 
		= \Lambda_R \times \Lambda_A \times \Lambda_B \ . 
		\label{eq:detector_decomposition_Lambda}
	\end{equation}
	
	We view $\Lambda_A$ and $\Lambda_B$ as the respective sample space for 
	either detector. As in a Bell model, it is implicitly assumed that both detectors 
	are of the same type; nonetheless, $\Lambda_A$ and $\Lambda_B$ need not be identical, 
	as other physical circumstances may lead to an asymmetry.  
	The set $\Lambda_R$ is the sample space for all other random 
	phenomena, including and especially for the particles emanating from 
	the source. It is worthy of note that $\Lambda_A$ and $\Lambda_B$ need not 
	only refer to ``intrinsic properties'' of the detector, but may very well refer 
	to how the detectors interact with the environment---including the particles. 
	
	Due to Eq. \eqref{eq:detector_decomposition_Lambda}, a 
	sample $\lambda \in \Lambda$ may be written as a triple 
	\begin{equation}
			\lambda = \left(\lambda_R,\lambda_A, \lambda_B\right)
		\label{eq:detector_decomposition_lambda}
	\end{equation}
	for some $\lambda_R \in \Lambda_R$ and $\lambda_A \in \Lambda_A$, and 
	$\lambda_B \in  \Lambda_B$. We call 
	$\lambda_A$ and 
	$\lambda_B$ \emph{random detector parameters} for 
	detector $A$ and $B$, respectively. 
	
	Eq. \eqref{eq:detector_decomposition_Lambda} constitutes an insubstantial 
	assumption on $\Lambda$: the ``splitting'' is physically motivated 
	and  
	we can always enlarge the sample space and assign probability zero to 
	events, which lie fully outside the original sample space. Furthermore, 
	$\Lambda_A$, $\Lambda_B$, or $\Lambda_R$ may in principle contain a single 
	element. 
	
	For the collection of random events 
	$\mathcal{A}$, we choose the product 
	\begin{equation}
				\mathcal{A} =
			\mathcal{A}_R \otimes \mathcal{A}_A \otimes  \mathcal{A}_B \ , 
			\label{eq:detector_decomposition_A}
	\end{equation}
	(cf. Def. 14.4 in \cite{klenkeProbabilityTheoryComprehensive2020}). 
	Of course, $\mathcal{A}$ may in principle be larger, but the choice in 
	Eq. \eqref{eq:detector_decomposition_A} is canonical. 
	
	If we use a single  
	probability measure $\mathbb{P}'$ (which does not depend on the detector 
	settings $a, b \in S$), then we may consider any corresponding Bell model 
	\begin{equation}
		\left(\Lambda_R \times \Lambda_A \times \Lambda_B, 
			\mathcal{A}_R \otimes \mathcal{A}_A \otimes  \mathcal{A}_B, 
			\mathbb{P}', S, A, B\right)
			\label{eq:noncontextualcontextualBell}
	\end{equation}
	and our general analysis in Sec. \ref{sec:discussion} applies. 
	In particular, if the Bell model is strongly local, then 
	the BCHSH inequality applies (cf. Cor. \ref{Cor:BCHSHcontra}). 

	However, in such a probabilistic model 
	the independence of $\mathbb{P}'$ from the detector settings implies that  
	a change of these settings does not change how the random detector parameters 
	$\lambda_A$ and $\lambda_B$ are distributed. Since they are explicitly 
	associated to the respective detectors, this constitutes an unphysical 
	assumption in general---even for ``classical'' Bell experiments. 
	Accordingly, we require more general 
	probabilistic models for Bell experiments in which the 
	(joint) probability measure may depend on the detector settings. 
	Due to this dependence on the experimental context  
	and the overt relationship to the related notion 
	of ``Kochen-Specker contextuality'' 
	in quantum mechanics \cite{shimonyExperimentalTestsLocal1971,
	kochenProblemHiddenVariables1975,
	budroniKochenSpeckerContextuality2022,
	spekkensContextualityPreparationsTransformations2005}, we shall call such 
	probabilistic models \emph{contextual}. Here de la Pe\~na, Cetto and Brody 
	\cite{delapenaHiddenvariableTheoriesBell1972}, 
	Lochak \cite{lochakHasBellsInequality1976}, as well as 
	Bohm and Hiley \cite{bohmNonlocalityQuantumTheory1981} are credited 
	for first pointing out this limitation of Bell models. 
	As noted before, other authors raised similar objections  
	\cite{pitowskyResolutionEinsteinPodolskyRosenBell1982,
	demuynckBellInequalitiesTheir1986,bransBellsTheoremDoes1988,
	szaboRealMeaningBells1994,
	feldmannNewLoopholeEinsteinPodolskyRosen1995,
	nagasawaLocalityHiddenvariableTheories1997,
	khrennikovContextualistViewpointGreenberger2001,
	khrennikovMathematicianViewpointBell2007,khrennikovViolationBellsInequality2009,
	nieuwenhuizenWhereBellWent2009,khrennikovProbabilityRandomnessQuantum2016,
	kupczynskiCanWeClose2017,
	kupczynskiEinsteinianNosignallingViolated2017,
	hanceBellsTheoremAllows2022,kupczynskiContextualityNonlocalityWhat2023,
	papatryfonosStaticBellTest2024,
	arroyoFamilyDeterministicModels2025,
	papatryfonosProposedExperimentsDetecting2025,
	lucWhatAreBearers2026,
	bacciagaluppiExtendingBellsTheorem2026}. 
				
	Accordingly, we shall generalize our model \eqref{eq:noncontextualcontextualBell} by 
	introducing an explicit dependence of the probability measure on both setting 
	parameters $a$ and $b$. 
	\begin{Definition}
		\label{Def:generalizedBM}
		A \emph{contextual Bell model} is a tuple 
		\begin{equation}
			\left(\Lambda_R,\mathcal{A}_R,\Lambda_A,\mathcal{A}_A,\Lambda_B,
			\mathcal{A}_B, \mathbb{P}_{\, . \,}, S, A, B\right)
			\label{eq:tuple_generalizedBM}
		\end{equation}
		such that the following holds: 
		\begin{enumerate}[1)]
			\item 	$(\Lambda_R,\mathcal{A}_R)$, $(\Lambda_A,\mathcal{A}_A)$, 
						and $(\Lambda_B,\mathcal{A}_B)$ are measurable spaces.   
			\item 	$S$ is a nonempty set. 
			\item 	$\Lambda$ and $\mathcal{A}$ are given via 
					Eqs. \eqref{eq:detector_decomposition_Lambda} 
					and \eqref{eq:detector_decomposition_A}, and 
					$\mathbb{P}_{\, . \,}$ is a family of probability measures such that 
					for every $a,b \in S$ the tuple 
					\begin{equation}
						 \left(\Lambda, 
						 \mathcal{A}, \mathbb{P}_{ab}\right) 
					\end{equation}
					is a probability space. 
			\item 	$A$ and $B$ are given via 
					Eqs. \eqref{eq:defAgen} and \eqref{eq:defBgen}, such that  
					for every $a,b \in S$ the functions 
					$A(a,b)$ and $B(a,b)$ are random variables 
					on $\left(\Lambda, \mathcal{A}, \mathbb{P}_{ab}\right)$. 
		\end{enumerate}
	\end{Definition}

	The above definition generalizes Def. \ref{Def:Bellmodel} with the 
	physically motivated restriction that $\Lambda$ and $\mathcal{A}$ are given via 
	Eqs. \eqref{eq:detector_decomposition_Lambda} 
	and \eqref{eq:detector_decomposition_A}. In this sense, 
	a contextual Bell model is a Bell model, if and only if the dependence of 
	the probability measure 
	$\mathbb{P}_{ab}$ on $a, b \in S$ is trivial. Accordingly, Bell models in 
	the sense of Def. \ref{Def:Bellmodel} 
	may be considered  
	``noncontextual''. 

	Contextual Bell models aim to describe   
	any Bell experiment with two similar detectors and a single source giving a binary  
	output: in a particular trial, the experimenter chooses the setting parameters 
	$a$ and $b$ and records the outcomes $A(a,b)(\lambda)$ and 
	$B(a,b)(\lambda)$. The central difference to ``noncontextual'' Bell 
	models is that a change of the setting parameters may also change 
	the probability that a particular $\lambda$ is realized. 
	
	We first consider a simple class of physically motivated examples. This 
	class corresponds to  
	the case that $\lambda_R$, $\lambda_A$, and $\lambda_B$ are  
	stochastically independent and their (marginal) distributions only depend 
	on the relevant setting parameter. 
	
	\begin{Example}
		\label{Ex:Pab_product}
	\begin{enumerate}[1)]
		\item 
			\label{itm:Ex:Pab_product1}
			Let $\mathbb{P}_R$ be a probability measure on the measurable space 
			$(\Lambda_R,\mathcal{A}_R)$ and let $S$ be a nonempty set. Furthermore, 
			let $\left(\mathbb{P}_{A,a} \right)_{a \in S}$ and 
			$\left( \mathbb{P}_{B,b} \right)_{b \in S}$ be families of 
			probability measures 
			on the measurable spaces 
			$(\Lambda_A,\mathcal{A}_A)$ and $(\Lambda_B,\mathcal{A}_B)$, 
			respectively. 
			
			Heuristically, the probability measure $\mathbb{P}_{A,a}$ 
			only depends on the setting $a$ of detector $A$ and describes the 
			distribution of 
			the random detector parameter $\lambda_A$ for each such choice. 
			The interpretation of 
			$\mathbb{P}_{B,b}$ for detector $B$ is analogous. The probability measure 
			$\mathbb{P}_{R}$ describes all other random phenomena and is 
			therefore independent 
			of both settings. 
		
			Now define $\Lambda$ and $\mathcal{A}$ via  
			Eqs. \eqref{eq:detector_decomposition_Lambda} and 
			\eqref{eq:detector_decomposition_A} 
			above. Then for all $a,b \in S$ the product measure 
				\begin{equation}
						\mathbb{P}_{ab} = 
					\mathbb{P}_{R} \otimes \mathbb{P}_{A,a} \otimes \mathbb{P}_{B,b}		
					\label{eq:Pab_product}
				\end{equation}
			indeed defines a probability measure on $\left(\Lambda, \mathcal{A}\right)$
			(cf. Ex. 14.15 in \cite{klenkeProbabilityTheoryComprehensive2020}). 
			It directly follows 
			that the
			probability measures $\mathbb{P}_R$, $\mathbb{P}_{A,a}$ and 
			$\mathbb{P}_{B,b}$ are the respective marginal distributions of 
			$\lambda_R$, $\lambda_A$, and 
			$\lambda_B$. By construction, $\lambda_R$, $\lambda_A$, and $\lambda_B$ 
			are stochastically independent: 
			for any $U_R \in \mathcal{A}_R$, $U_A \in \mathcal{A}_A$, and 
			$U_B \in \mathcal{A}_B$ it holds that 
			\begin{multline}
				\mathbb{P}_{ab}\left(U_R \times U_A \times U_B \right)
				= 
				\\ 
				\mathbb{P}_{ab}\left(U_R \times \Lambda_A \times \Lambda_B \right) \, 
				\mathbb{P}_{ab}\left(\Lambda_R \times U_A \times \Lambda_B \right) \, 
				\mathbb{P}_{ab}\left(\Lambda_R \times \Lambda_A \times U_B \right) \ . 
			\end{multline}
		
			By choosing any $A$ and $B$ as in Def. \ref{Def:generalizedBM}, 
			we hence obtain a contextual Bell model.
		\item 
			\label{itm:Ex:Pab_product2}
			As an illustrative example, we consider a special case 
			of \ref{itm:Ex:Pab_product1}, 
			using a fair coin and two fair 
			dice with $100$ sides. 
			
			For $p \in [0,1]$, denote 
			by $\operatorname{Rad}_{p}$	the Rademacher distribution on 
			$\lbrace -1, +1 \rbrace$ together with 
			the power set $2^{\lbrace -1, +1 \rbrace}$. Its values are given via 
			\begin{subequations}
			\begin{align}
				\operatorname{Rad}_{p}(\lbrace +1 \rbrace) &=
				\operatorname{Rad}_{p}(+1) =p	
				\\
				\operatorname{Rad}_{p}(\lbrace -1 \rbrace) &= 
				\operatorname{Rad}_{p}(-1 ) =1-p			
			\end{align}
			\end{subequations}
			(cf. Ex. 1.105(i) in \cite{klenkeProbabilityTheoryComprehensive2020}). 
			
			For the coin flip, we use $\mathbb{P}_R = \operatorname{Rad}_{1/2}$. 
			For the first die, we choose a setting 
			$a \in S=\lbrace 1, 2, \dots, 100 \rbrace \subset \N$. 
			Then $\lambda_A=-1$, if the outcome is below or equal to 
			$a$ and $\lambda_A=+1$, if the 
			outcome is above $a$. Thus, 
			$\mathbb{P}_{A,a} = \operatorname{Rad}_{1-a/100}$. Analogously, for the 
			second die we choose $\mathbb{P}_{B,b} = \operatorname{Rad}_{1-b/100}$. 
			
			The detector random variables we define via simple multiplication: 
			\begin{equation}
				A(a,b)(\lambda_R,\lambda_A) = \lambda_R \lambda_A
				\quad \text{and} \quad 
				B(a,b)(\lambda_R,\lambda_B) = - \lambda_R \lambda_B
				\ .
			\end{equation}
			Recalling Eq. \eqref{eq:defpjoint}, 
			a direct computation reveals that 
			\begin{equation}
				p_{\pm \pm}(a,b) = 
				\left( \operatorname{Rad}_{a/100}(\mp 1)
				\, \operatorname{Rad}_{b/100}(\pm 1) +
				\operatorname{Rad}_{a/100}(\pm 1)
				\, \operatorname{Rad}_{b/100}(\mp 1) 
				 \right) / 2 \ . 
			\end{equation}
			
			It is a notable property of the Bell experiment 
			described by this contextual Bell model  
			that we can provide a ``physically equivalent'' description via a 
			``noncontextual'' Bell model. This is achieved 
			by taking $\lambda_A$ and $\lambda_B$ to be 
			the outcomes of the die rolls instead, so that the dependence on 
			the settings is shifted to the random detector variables. 
	\end{enumerate} 
	\end{Example}	
	
	\begin{Remark}
		\label{Rem:contextuality}
	While Defs. \ref{Def:Bellmodel} and \ref{Def:generalizedBM} describe  
	mathematically distinct classes of probabilistic models, 
	Ex. \ref{Ex:Pab_product}.\ref{itm:Ex:Pab_product2} 
	shows that there exist Bell experiments that can be described by either. 
	Commonly, an appropriate ``change of variables'' relates the 
	contextual to the ``noncontextual'' case. Still, as we will see in 
	Sec. \ref{sec:poc_example} below, there do exist contextual Bell models 
	that cannot be reduced to the ``noncontextual'' case while retaining 
	certain properties of ``locality''. 
	The question under which conditions 
	a contextual Bell model can be ``equivalently described'' via a 
	``noncontextual'' Bell model appears to be 
	a worthy subject of further inquiry. Both this question as well as the 
	complementary one of what 
	``contextuality'' is in a mathematically well-defined sense
	are, however, beyond the scope of this work. 
	\end{Remark}
	
	Of course, not all contextual Bell models rely on 
	probability measures of the type \eqref{eq:Pab_product}. 
	For a general description of physically realistic Bell experiments, 
	Ex. \ref{Ex:Pab_product} is too restrictive: 
	$\lambda_R$, $\lambda_A$, and $\lambda_B$ may be stochastically dependent 
	on each other. Still, this raises the question of how to choose the measure 
	$\mathbb{P}_{\, . \,}$, so that 
	we can nonetheless affiliate the distribution of $\lambda_A$ and 
	$\lambda_B$ to the setting of the respective detector. 
	
	We shall employ this question as a motivation for the introduction 
	of ``strong'' and ``weak locality'' for 
	contextual Bell models. This is in analogy to our analysis of the 
	``noncontextual'' case 
	in Sec. \ref{sec:discussion}. 
	
	We first observe that Def. \ref{Def:generalizedBM} 
	does not introduce any explicit or implicit 
	assumption of ``locality''. In this respect the definition 
	follows the definition of a ``noncontextual'' Bell model 
	in Def. \ref{Def:Bellmodel}. This is the 
	reason why there is not any explicit affiliation 
	of (the distribution of) $\lambda_A$ and 
	$\lambda_B$ to the detector settings. In principle, 
	the setting of detector $A$ could determine the probability 
	that a certain $\lambda_B$ is realized and vice versa. As long as 
	one does not impose any additional ``locality'' constraint, a contextual 
	Bell model can describe such a Bell experiment. 
	
	In Ex. \ref{Ex:Pab_product} 
	we implicitly made such a locality assumption via our choice 
	of the marginal distributions. Though in general Eq. \eqref{eq:Pab_product} is 
	too restrictive, in a ``strongly local'' contextual Bell model we 
	may still expect (families of) probability measures 
	$\mathbb{P}_R$, $\mathbb{P}_{A,a}$, and $\mathbb{P}_{B,b}$ 
	as in Ex. \ref{Ex:Pab_product} to be obtained as 
	marginal distributions of $\mathbb{P}_{ab}$. 
	Specifically, if $\pr_R$, $\pr_A$, and $\pr_B$ denote 
	the respective projections, then the following identities ought to hold:  
	\begin{subequations}
		\label{eq:conditions_margin}
	\begin{align}
		\mathbb{P}_{R} &= \mathbb{P}_{ab} \circ \pr_R^{-1}
		\label{eq:conditions_marginR}
		\\
		\mathbb{P}_{A,a}
		&= 
		\mathbb{P}_{ab} \circ \pr_A^{-1}
		\label{eq:conditions_marginA}
		\\
		\mathbb{P}_{B,b}
		&= 
		\mathbb{P}_{ab} \circ \pr_B^{-1}
		\label{eq:conditions_marginB}
	\end{align}
	\end{subequations}
	The respective independence of the above marginal 
	distributions on the parameters $a$ and $b$ is then supposed to be 
	a consequence of ``strong locality''. 
	
	In our motivation of a suitable such notion, 
	we may further construct examples of joint probability measures $\mathbb{P}_{ab}$, 
	which satisfy Eqs. \eqref{eq:conditions_margin} and generalize 
	Eq. \eqref{eq:Pab_product}. 
	
	To this end, we 
	consider the specific case that detector $A$ with setting $a$ may register 
	the choice of $\lambda_R$ via ``local interactions'' and choose 
	$\lambda_A$ accordingly. Analogously, detector $B$ may choose $\lambda_B$ 
	upon registering $\lambda_R$. Even a strict reading of the physical 
	requirement of ``locality'' ought to allow for such behavior. 
	
	Mathematically, we use stochastic kernels for this purpose. 
	In the literature, they are also known as Markov kernels, see e.g. 
	Def. 8.25 in \cite{klenkeProbabilityTheoryComprehensive2020}. For 
	each $a$, $b \in S$ we introduce the stochastic kernels 
	\begin{subequations}
		\label{eq:stochkernel}
	\begin{align}
		\tilde{\mathbb{P}}_{A,a} \colon \Lambda_R \times \mathcal{A}_A \to [0,1] 
		\ , \quad (\lambda_R,U) \mapsto \tilde{\mathbb{P}}_{A,a} (\lambda_R,U)
		\\
		\tilde{\mathbb{P}}_{B,b} \colon \Lambda_R \times \mathcal{A}_B \to [0,1]
		\ , \quad (\lambda_R,V) \mapsto \tilde{\mathbb{P}}_{B,b} (\lambda_R,V)
	\end{align}
	\end{subequations}
	such that for any $U \in \mathcal{A}_A$ and $V \in \mathcal{A}_B$ we have 
	the following: 
	\begin{subequations}
	\begin{align}		
			\mathbb{P}_{A,a}(U) &= \int_{\Lambda_R} \d \mathbb{P}_R(\lambda_R)
			\int_U \d \tilde{\mathbb{P}}_{A,a}(\lambda_R, \lambda_A) 
				\\
			\mathbb{P}_{B,b}(V) &= \int_{\Lambda_R} \d \mathbb{P}_R(\lambda_R)
			\int_V \d \tilde{\mathbb{P}}_{B,b}(\lambda_R, \lambda_B) 
	\end{align}
	\end{subequations}
	Formally, we write the joint probability measure $\mathbb{P}_{ab}$ 
	as 
	\begin{equation}
			\mathbb{P}_{ab} = 
		\mathbb{P}_{R} \otimes \tilde{\mathbb{P}}_{A,a} \otimes \tilde{\mathbb{P}}_{B,b}	
		\ .	
		\label{eq:Pab_kernel}
	\end{equation}
	This defines the probability that $\lambda = (\lambda_R, \lambda_A, \lambda_B)$ 
	takes values in $U \in \mathcal{A}$ to be  
	\begin{equation}
		\mathbb{P}_{ab}(U) 
			= 
			\int_{U} \d \mathbb{P}_R(\lambda_R) \, 
			\d \tilde{\mathbb{P}}_{A,a}(\lambda_R, \lambda_A) \, 
			\d \tilde{\mathbb{P}}_{B,b}(\lambda_R, \lambda_B) \ .
	\end{equation}
	\begin{Lemma}
		\label{Lem:Pab_kernel_probspace}
		$\left(\Lambda,\mathcal{A},\mathbb{P}_{ab} \right)$, as defined via 
		Eqs. 
		\eqref{eq:detector_decomposition_Lambda}, \eqref{eq:detector_decomposition_A}, 
		and \eqref{eq:Pab_kernel}
		is a probability space for every ${a,b \in S}$. 
		Furthermore, there exist families of probability measures $\mathbb{P}_R$, 
		$(\mathbb{P}_{A,a})_{a \in S}$, and $(\mathbb{P}_{B,b})_{b \in S}$ such that Eqs. 
		\eqref{eq:conditions_margin} hold for all ${a,b \in S}$. 
	\end{Lemma}
	\begin{Proof}
		By Cor. 14.26 in \cite{klenkeProbabilityTheoryComprehensive2020}, 
		$\mathbb{P}_{R} \otimes \tilde{\mathbb{P}}_{A,a}$ is a probability measure 
		on $(\Lambda_R \times \Lambda_A, \mathcal{A}_R \times \mathcal{A}_A)$. 
		Since we may view  $\tilde{\mathbb{P}}_{B,b}$ as a transition kernel from 
		$\Lambda_R \times \Lambda_A$ to $\Lambda_B$ with trivial dependence on 
		$\lambda_A$, reapplication of the corollary 
		yields the first assertion. The second assertion follows from a direct computation 
		of the marginals. 
	\end{Proof}	
	
	The fact, that the measure in Eq. \eqref{eq:Pab_kernel} indeed generalizes 
	the one in 
	Eq. \eqref{eq:Pab_product}, can be shown by setting 
	\begin{equation}
		\tilde{\mathbb{P}}_{A,a} (\lambda_R,U) = \mathbb{P}_{A,a} (U) 
		\quad \text{and} \quad 
		\tilde{\mathbb{P}}_{B,b} (\lambda_R,V)  = \mathbb{P}_{B,b} (V)
	\end{equation}
	for any $\lambda_R \in \Lambda_R$, $U \in \mathcal{A}_A$, $V \in \mathcal{A}_B$, 
	and $a,b \in S$. 
	
	We now use the above construction to provide an explicit, physical example 
	of a ``strongly local'' contextual Bell model. As a first step, 
	we construct the setting-dependent probability measure. 
	
	\begin{Example}
		\label{Ex:correlatedP}
		We consider a ``classical'' Bell experiment, where two different masses 
		are sent to detector 
		$A$ and $B$ with some given speed, respectively. The mass going to 
		detector $A$ is $(3+\lambda_R)/2$ and the mass going to detector $B$ 
		is $(3-\lambda_R)/2$ (in terms of any mass unit) with 
		$\lambda_R \in \lbrace -1, +1 \rbrace = \Lambda_R$. 
		The probability for each $\lambda_R$ is $1/2$. Hence, $\mathbb{P}_R$ is 
		the Rademacher distribution $\operatorname{Rad}_{1/2}$ on $\Lambda_R$
		and $\mathcal{A}_R$ is the power set $2^{\Lambda_R}$ 
		(cf. Ex. \ref{Ex:Pab_product}.\ref{itm:Ex:Pab_product2}). 
		
		Once the incoming mass arrives at detector $A$, the detector measures it and 
		then places a second mass of size $\lambda_A \in (0,\infty) = \Lambda_A$ 
		at rest in front of the incoming 
		mass. The distribution for this second mass depends on the 
		detector setting $a \in (1,\infty)=S$ and the incoming mass, so that 
		$\lambda_A$ stochastically depends on $\lambda_R$. Specifically, 
		the mass $\lambda_A$ follows an exponential distribution with 
		mean $(a+\lambda_R)$. Afterwards, the two masses unite in 
		a perfectly inelastic collision. The detector checks, if the total mass 
		meets a certain 
		threshold speed or not, yielding a respective 
		value of $+1$ and $-1$. Detector $B$ 
		behaves analogously with $\Lambda_B = (0,\infty)$ and $b \in S$. 
		For the classes of events we choose the corresponding Lebesgue sets. 
		
		On 
		\begin{equation}
				\Lambda = \lbrace \pm 1 \rbrace \times (0,\infty)^2 
		\end{equation}
		we therefore consider the following probability density function 
		\begin{equation}
			\rho_{ab}(\lambda_R,\lambda_A,\lambda_B) = 
			\frac{1}{2 (a+\lambda_R) (b-\lambda_R)}  \, 
				e^{- \lambda_A / (a+\lambda_R) - \lambda_B / (b-\lambda_R)}
			\label{eq:example_rhoab}
		\end{equation}
		for the measure $\mathbb{P}_{ab}$. The stochastic kernels 
		$\tilde{\mathbb{P}}_{A,a}$ and $\tilde{\mathbb{P}}_{B,b}$ 
		as well as the marginal distributions 
		$\mathbb{P}_R$, $\mathbb{P}_{A,a}$, 
		$\mathbb{P}_{B,b}$ are then obtained via integration/summation. 
	\end{Example}

	In the next step, we ask what conditions ``strong locality'' ought to impose 
	on the 
	(families of) 
	detector random variables $A$ and $B$. Of course, 
	different choices of 
	$b$ and $\lambda_B$ ought to yield the same value of $A(a,b)(\lambda)$. 
	The analogous property should hold for the family of random variables $B$. 	
	A practical illustration of this requirement is provided by the following 
	continuation of Ex. \ref{Ex:correlatedP} above. 
	
	\begin{Example}
		\label{Ex:correlatedP_cont}
		We set the initial 
		speed of either mass to unity, denote the threshold speed of either detector 
		by $\xi \in [0, 1]$, and write $\theta$ for the Heaviside function. 
		
		According to the law of inelastic collision for point masses, 
		suitable detector random variables $A$ and $B$ on $\Lambda$ are given as 
		follows:  
		\begin{subequations}
			\label{eq:defAB_correlatedP_cont}
		\begin{align}
				A(\lambda_R, \lambda_A, \lambda_B) &=
				2 \, \theta \left( 
					\frac{3-\lambda_R}{2 \lambda_A + 3 - \lambda_R} - \xi 
					\right) - 1
			\\
			B(\lambda_R, \lambda_A, \lambda_B) &= 
				2 \, \theta \left( 
					\frac{3+\lambda_R}{2 \lambda_B + 3 + \lambda_R} - \xi 
					\right) - 1
		\end{align}
		\end{subequations}
	\end{Example}

	A noteworthy observation in Ex. \ref{Ex:correlatedP_cont} as well 
	as previously in Ex. \ref{Ex:Pab_product}.\ref{itm:Ex:Pab_product2}
	is that 
	the detector random variables $A$ and $B$ 
	do not need to explicitly depend on 
	the respective setting parameters: the (marginal) 
	distributions of $\lambda_A$ and $\lambda_B$ will generally exhibit such a 
	dependence already. Nonetheless, we may allow for such an explicit dependence, 
	provided that the same parameter value is used for the detector random variable. 
	
	Before we abstract a definition of ``strong locality'' for 
	contextual Bell models from our discussion, we shall 
	consider a suitable definition of ``weak locality''.  
	Here we may follow the respective definition for Bell models,
	Def. \ref{Def:locality}: a contextual Bell model is ``weakly local'', if 
	the distribution of $A(a,b)$ depends on $a$, but not on $b$, and the distribution of 
	$B(a,b)$ depends on $b$, but not on $a$. 
	
	This loose definition shows, however, that the conditions for 
	``strong locality'' provided so far do not suffice to imply 
	``weak locality''. To see this, denote by $\pr_{RA}$ and $\pr_{RB}$ the 
	projections of $\Lambda$ to the factors $\Lambda_R \times \Lambda_A$ and 
	$\Lambda_R \times \Lambda_B$, respectively. Then the marginal distributions 
	$\mathbb{P}_{ab} \circ \pr_{RA}^{-1}$ and 
	$\mathbb{P}_{ab} \circ \pr_{RB}^{-1}$ may in principle depend on both 
	detector settings $a$ and $b$, even if Eqs. \eqref{eq:conditions_margin} 
	are satisfied. Loosely speaking, the 
	``off-diagonal probabilities'' may depend on the other setting parameter, even 
	if the ``diagonals'' do not. In that case ``weak locality'' may be violated. 
	Yet we would like to assure that it is indeed implied. 
	
	Accordingly, we arrive at the following formal definition. 
	\begin{Definition}
		\label{Def:generalizedBM_locality} 
		Consider a contextual Bell model as in Def. 
		\ref{Def:generalizedBM}. 
		Denote by $\pr_R$, $\pr_{RA}$, and $\pr_{RB}$ the projections of 
		$\Lambda$ to the factors $\Lambda_R$, 
		$\Lambda_R \times \Lambda_A$, and $\Lambda_R \times \Lambda_B$, respectively.
		\begin{enumerate}[1)]
			\item 
			\label{itm:generalizedBM_stronglylocal}
			The contextual Bell model is called \emph{strongly local}, if 
					all of the following hold: 
				\begin{enumerate}[a)]	
					\item				
						\label{itm:generalizedBM_stronglylocal1}
						For all $a,a',b,b' \in S$ the respective marginal 
						distributions satisfy 
						\begin{subequations}
						\label{eq:generalizedBM_stronglylocal1}
						\begin{align}
							\mathbb{P}_{ab} \circ \pr_{RA}^{-1}
							&= 
							\mathbb{P}_{ab'} \circ \pr_{RA}^{-1} \ ,
							\label{eq:generalizedBM_stronglylocal1_b}
							\\
							\mathbb{P}_{ab} \circ \pr_{RB}^{-1}
							&= 
							\mathbb{P}_{a'b} \circ \pr_{RB}^{-1} \ . 
						\end{align}
						\end{subequations}
					\item 	
						\label{itm:generalizedBM_stronglylocal2}
					For all $a,a',b,b' \in S$, and 
							almost all (with respect to $\mathbb{P}_{ab}$) 
							$\lambda_R \in \Lambda_R$, 
							$\lambda_A,\lambda_A' \in \Lambda_A$, 
							and $\lambda_B,\lambda_B' \in \Lambda_B$ 
							it holds that 
						\begin{subequations}
						\label{eq:generalizedBM_stronglylocal2} 
						\begin{align}
							A (a,b)(\lambda_R,\lambda_A, \lambda_B ) &= 
								A(a,b') (\lambda_R,\lambda_A, \lambda_B ) \ ,
							\label{eq:generalizedBM_stronglylocal2_a1}
							\\
							B(a,b)(\lambda_R,\lambda_A, \lambda_B ) &= 
								B(a',b)(\lambda_R,\lambda_A, \lambda_B ) \ ,
							\label{eq:generalizedBM_stronglylocal2_b1}
							\\
							A (a,b)(\lambda_R,\lambda_A, \lambda_B ) &= 
								A(a,b) (\lambda_R,\lambda_A, \lambda_B' ) \ ,
							\label{eq:generalizedBM_stronglylocal2_a2}
							\\
							B(a,b)(\lambda_R,\lambda_A, \lambda_B ) &= 
							B(a,b)(\lambda_R,\lambda_A', \lambda_B ) \ .
						\end{align}
						\end{subequations}			
				\end{enumerate} 
			\item 
				\label{itm:generalizedBM_weaklylocal}
				The contextual Bell model is called \emph{weakly local}, if 
					for all $a, a', b, b' \in S$ 
					the respective probability distributions satisfy
					\begin{subequations}
					 	\label{eq:generalizedBM_weaklylocal}
					\begin{align}
						\mathbb{P}_{ab} \circ \bigl(A(a,b) \bigr)^{-1} & = 
							\mathbb{P}_{ab'} \circ \bigl(A(a,b') \bigr)^{-1} 
							\ , 
							\\ 
						\mathbb{P}_{ab} \circ \bigl(B(a,b) \bigr)^{-1} &= 
							\mathbb{P}_{a'b} \circ \bigl(B(a',b) \bigr)^{-1} \ . 
					\end{align}						
					\end{subequations}	
		\end{enumerate}	
	\end{Definition}
	
	In simpler terms, Def. \ref{Def:generalizedBM_locality} states that 
	a contextual Bell model is strongly local, if both of the following hold: 
	\begin{enumerate}[a)]	
		\item	\label{itm:stronglclgenBell1}
				The  
				marginal distributions of $(\lambda_R, \lambda_A)$ 
				as well as $(\lambda_R, \lambda_B)$ do not depend on $b$ and $a$, 
				respectively, and the marginal distribution 
				of $\lambda_R$ does not depend on either parameter. 
		\item 	For each $a, b \in S$ the detector random variables 
				may ($\mathbb{P}_{ab}$-almost surely) be written as functions 
				\begin{subequations}
					\label{eq:defAB_detector}
				\begin{align}
					A(a)\colon 
						\Lambda \to \lbrace -1, +1\rbrace \ , \quad
						(\lambda_R,\lambda_A,\lambda_B) 
						\mapsto A(a)(\lambda_R,\lambda_A)
						\label{eq:defA_detector}
						\\
					B(b)\colon 
						\Lambda \to \lbrace -1, +1\rbrace \ , \quad
						(\lambda_R,\lambda_A,\lambda_B) 
						\mapsto B(b)(\lambda_R,\lambda_B)
						\label{eq:defB_detector} 
				\end{align}
				\end{subequations}
	\end{enumerate}
	It is worth noting that the above notion of strong locality is implicit in a 
	previous expression 
	by Nieuwenhuizen (cf. Eq. 5 in \cite{nieuwenhuizenWhereBellWent2009}). 
	
	Property \ref{itm:stronglclgenBell1} above implies that 
	there exist marginal distributions 
	$\mathbb{P}_{A,a}$ and $\mathbb{P}_{B,b}$ depending on 
	$a$ and $b$, respectively, such that 
	Eqs. \eqref{eq:conditions_marginA} and \eqref{eq:conditions_marginB} hold. 
	It furthermore implies 
	that there exists a marginal distribution  
	$\mathbb{P}_R$ independent of $a$ and $b$ such that 
	Eq. \eqref{eq:conditions_marginR} holds. 
	The definition of weak locality in Def. 
	\ref{Def:generalizedBM_locality} agrees with the loose 
	definition provided before. 
	
	It remains to show that strong locality indeed implies weak locality. 
	\begin{Lemma}
		\label{Lem:genBM_strong_implies_weak}
		Any strongly local contextual Bell model is weakly local. 
	\end{Lemma}
	\begin{Proof}
		We only consider $A$; the argument for $B$ is analogous. 
		By Eqs. \eqref{eq:generalizedBM_stronglylocal2_a1} and 
		\eqref{eq:generalizedBM_stronglylocal2_a2} there exist a function 
		$\tilde{A}$, such that for all $a,b \in \R$ 
		\begin{equation}
			A(a,b) = \tilde{A}(a) \circ \pr_{RA} 
		\end{equation}
		($\mathbb{P}_{ab}$-almost surely). 
		$\tilde{A}(a)$ is measurable by definition of the product 
		$\sigma$-algebra and because 
		$A(a,b)$ is measurable.  
		Thus 
		\begin{equation}
				\mathbb{P}_{ab} \circ \bigl(A(a,b) \bigr)^{-1} = 
				\left( \mathbb{P}_{ab} \circ \pr_{RA}^{-1} \right) \circ 
				\tilde{A}(a)^{-1} \ . 
		\end{equation}
		By Eq. \eqref{eq:generalizedBM_stronglylocal1_b} this only depends on $a$.  
	\end{Proof}
	
	We shall also show that the two notions of locality for 
	contextual Bell models are consistent with those for 
	``noncontextual'' Bell models. 
	\begin{Lemma}
		\label{Lem:genBMlocality_consistent}
		Given a contextual Bell model as in Def. \ref{Def:generalizedBM}, suppose 
		that there exists a probability measure $\mathbb{P}'$ on 
		$\left(\Lambda, \mathcal{A}\right)$ such that 
		$\mathbb{P}' = \mathbb{P}_{ab}$ for all $a,b \in S$. 
		
		Then the tuple \eqref{eq:noncontextualcontextualBell} 
		is a Bell model (Def. \ref{Def:Bellmodel}) and the following holds: 
		\begin{enumerate}[1)]
			\item If the contextual Bell model is strongly local, then 
					the Bell model \eqref{eq:noncontextualcontextualBell} 
					is strongly local. 
			\item If the contextual Bell model is weakly local, then 
					the Bell model \eqref{eq:noncontextualcontextualBell} 
					is weakly local.
		\end{enumerate}
	\end{Lemma}
	\begin{Proof}
		\begin{enumerate}[1)]
			\item This is trivial due to Eqs. 
			\eqref{eq:generalizedBM_stronglylocal2_a1} and 
			\eqref{eq:generalizedBM_stronglylocal2_b1}. 
			\item This holds by construction. 
		\end{enumerate}
	\end{Proof}
	
	The following proposition provides a means to construct examples of 
	strongly local contextual Bell models 
	from Lem. \ref{Lem:Pab_kernel_probspace}. An explicit physical example 
	is hence given by Exs. \ref{Ex:correlatedP} and \ref{Ex:correlatedP_cont}. 
	
	\begin{Proposition}
			\label{Prop:stoch_kernel_is_strongly_local}
		Let $\Lambda$, $\mathcal{A}$ and $S$ be given as in Def. 
		\ref{Def:generalizedBM}. For all 
		$a, b \in S$ let $\tilde{\mathbb{P}}_{A,a}$ and 
		$\tilde{\mathbb{P}}_{B,b}$ be stochastic kernels as in Eq. 
		\eqref{eq:stochkernel}. For every 
		$a, b \in S$ define $\mathbb{P}_{ab}$ 
		via Eq. \eqref{eq:Pab_kernel}. 
		Furthermore, for every 
		$a, b \in S$ let $A(a)$ and $B(b)$  
		be random variables on 
		the probability space 
		$\left(\Lambda,\mathcal{A},\mathbb{P}_{ab} \right)$ such 
		that Eqs. \eqref{eq:defAB_detector} hold. 
		
		By viewing $A$ and $B$ as functions on 
		$S \times S \times \Lambda$, 
		we obtain a strongly local contextual Bell model.
	\end{Proposition}	
	\begin{Proof}
		We have a contextual Bell model by construction in conjunction with 
		Lem. \ref{Lem:Pab_kernel_probspace}. The respective marginal distributions 
		are $\mathbb{P}_R \otimes \tilde{\mathbb{P}}_{A,a}$ 
		and $\mathbb{P}_R \otimes \tilde{\mathbb{P}}_{B,b}$. Hence, 
		Eqs. \eqref{eq:generalizedBM_stronglylocal1} hold. Eqs. 
		\eqref{eq:generalizedBM_stronglylocal2} also hold by construction. 
	\end{Proof}
		
	The following theorem shows that strongly local contextual Bell 
	models falling under the class given by 
	Prop. \ref{Prop:stoch_kernel_is_strongly_local} satisfy 
	the BCHSH inequality.
	
\begin{Theorem}
	\label{Thm:BCHSH_detector} 
	Under the hypotheses of 
	Prop. \ref{Prop:stoch_kernel_is_strongly_local}, for $a,b \in S$ 
	define the expectation value 
				\begin{equation}
					\Evalue(a,b) = \int_{\Lambda} \d \mathbb{P}_{ab} \, 
						 A(a) B(b) \ . 
					\label{eq:Eabsplit}
				\end{equation} 
				
	Then $\Evalue$ satisfies the BCHSH inequality \eqref{eq:BCHSH}.
\end{Theorem}
\begin{Proof} 
	The main tool of proof is Fubini's theorem for transition kernels 
	(cf. Thm. 14.32 in \cite{klenkeProbabilityTheoryComprehensive2020}). 
	
	Due to Eqs. \eqref{eq:generalizedBM_stronglylocal2}, we 
	drop the obsolete dependencies of $A(a)$ on $\lambda_B$ and of $B(b)$ 
	on $\lambda_A$. Then define the 
	random variables $\tilde{A}$ and $\tilde{B}$ 
	on the probability space 
	$\left(\Lambda_R,\mathcal{A}_R,\mathbb{P}_R \right)$ 
	via integration:  
		\begin{subequations}
		\begin{gather}
				\tilde{A}(a)(\lambda_R) = \int_{\Lambda_A} A(a)(\lambda_R,\lambda_A) \, 
					\d \tilde{\mathbb{P}}_{A,a}(\lambda_R,\lambda_A)
				\\
				\tilde{B}(b) (\lambda_R) = \int_{\Lambda_B} B(b)(\lambda_R,\lambda_B) \, 
					\d \tilde{\mathbb{P}}_{B,a}(\lambda_R,\lambda_B)
		\end{gather}
		\end{subequations}
	By construction, $\tilde{A}$ and $\tilde{B}$ take values in the interval 
	$[-1,1]$. 
	
	We now apply Fubini's theorem twice and use the above definitions:
	\begin{align}
			\Evalue(a,b) 
		&= 
			\int_\Lambda 
				\d \bigl( \mathbb{P}_{R} \otimes \tilde{\mathbb{P}}_{A,a} 
				\otimes \tilde{\mathbb{P}}_{B,b} \bigr) 
				\, A(a) \, B(b)
		\\ 
		&= 
			\int_{\Lambda_R \times \Lambda_A} 
				\d \bigl( \mathbb{P}_{R} \otimes \tilde{\mathbb{P}}_{A,a} 
				\bigr) (\lambda_R, \lambda_A)
				\int_{\Lambda_B}  
				\, A(a)(\lambda_R,\lambda_A) \, B(b)(\lambda_R,\lambda_B) 
				\, \d \tilde{\mathbb{P}}_{B,b} (\lambda_R,\lambda_B) 
		\\
		&= \int_{\Lambda_R \times \Lambda_A}
				\d \bigl( \mathbb{P}_{R} \otimes \tilde{\mathbb{P}}_{A,a} 
				\bigr) (\lambda_R, \lambda_A) \, 
				A(a)(\lambda_R,\lambda_A) \, \tilde{B}(b)(\lambda_R) 
		\\  
		&= \int_{\Lambda_R} 
				\d \mathbb{P}_{R} (\lambda_R) 
				\int_{\Lambda_A} \d \tilde{\mathbb{P}}_{A,a} 
				(\lambda_R, \lambda_A) 
				\, A(a)(\lambda_R,\lambda_A) \, \tilde{B}(b)(\lambda_R) 
	\end{align}
	Therefore, 
	\begin{equation}
		\label{eq:E_average}
		\Evalue(a,b) = \int_{\Lambda_R} \d \mathbb{P}_{R} \, 
			 \tilde{A}(a) \, \tilde{B}(b) \ . 
	\end{equation}
	Due to Rem. \ref{Rem:BCHSHothervalues}, the BCHSH inequality \eqref{eq:BCHSH} 
	is satisfied. 
\end{Proof}

Eq. \eqref{eq:E_average} suggests the following interpretation: 
the BCHSH inequality holds for this class of contextual 
Bell models, because the averages of $A(a)$ and $B(b)$ with regards to the 
respective random detector parameters yields a strongly local Bell model. 

Thm. \ref{Thm:BCHSH_detector} generalizes a previous result by 
Gill and Lambare \cite{gillKupczynskisContextualLocally2023}, for which 
$\lambda_R$, $\lambda_A$, and $\lambda_B$ were taken to be stochastically 
independent (cf. Eq. \eqref{eq:Pab_product} above). In particular, the class of models advocated for 
in \cite{kupczynskiEinsteinianNosignallingViolated2017} 
satisfy the BCHSH inequality (cf. Eq. 4 therein; see also 
\cite{gillGeneralCommentaryMoon2022}). 

The class 
of models of Thm. \ref{Thm:BCHSH_detector} still does not exhaust  
all possible strongly local contextual Bell models. This raises the question, 
whether there 
are probabilistic models of the latter kind that can violate the BCHSH inequality. 

The following adaption of Lem. \ref{Lem:uncorrelated+weaklylocal} 
provides a necessary condition for such a violation. 
\begin{Lemma}
	\label{Lem:uncorrelated+weaklylocal+contextual}
	Given a weakly local contextual Bell model as in 
	Defs. \ref{Def:generalizedBM} and \ref{Def:generalizedBM_locality}, 
	for $a,b \in S$ define $\Evalue$ via 
			\begin{equation}
				\Evalue(a,b) = \int_{\Lambda} \d \mathbb{P}_{ab} \, 
					 A(a,b) \, B(a,b) \ . 
				\label{eq:Eabsplit_full}
			\end{equation} 
	
	If $A(a,b)$ and $B(a,b)$ are uncorrelated with 
	respect to $\mathbb{P}_{ab}$ for all $a$ and $b$, then 
	the BCHSH inequality \eqref{eq:BCHSH} holds. 
\end{Lemma}
\begin{Proof}
	The proof is fully analogous to the one of Lem. 
	\ref{Lem:uncorrelated+weaklylocal}. 
\end{Proof}

Still, a correlation of the detector 
random variables for certain choices of settings 
does not necessarily imply a violation of the 
BCHSH inequality. This is shown by the 
following example. 

\begin{Example}
	\label{Ex:Gumbel}
We adapt Exs. \ref{Ex:correlatedP} and \ref{Ex:correlatedP_cont} 
	by replacing the family of joint 
	probability measures $(\mathbb{P}_{ab})_{a,b \in S}$.  
	
	For this purpose, we use a 
	distribution by Gumbel \cite{gumbelBivariateExponentialDistributions1960} 
	for two positively correlated, exponentially distributed random 
	variables. 
	Under a suitable choice of parameters and for 
	$x,y \in [0,\infty)$, this distribution 
	is given by the probability density function $f$ with 
	\begin{equation}
		f(x,y) = e^{-x-y-xy} \bigl( (1+x) (1+y) - 1 \bigr) 
	\end{equation}
	(cf. Eq. 2.3 in \cite{gumbelBivariateExponentialDistributions1960}). 
	
	As in Ex. \ref{Ex:correlatedP}, 
	$\mathbb{P}_{ab}$ is defined via a probability density function $\rho_{ab}$: 
		\begin{equation}
			\rho_{ab}(\lambda_R,\lambda_A,\lambda_B) = 
			\frac{f\bigr(\lambda_A /(a+\lambda_R), \lambda_B / (b-\lambda_R)\bigl)}
			{2 (a+\lambda_R) (b-\lambda_R)} \ . 
			\label{eq:example_rhoab_correlated}
		\end{equation}
	Note that this choice is mathematically motivated; we do not 
	provide any physical reason for it. 

	Again, we consider $A$ and $B$ as given via Eq. 
	\eqref{eq:defAB_correlatedP_cont}. 
	In order to show that they are correlated, we first define 
	\begin{equation}
		p_{\pm \mp}(a,b) = \mathbb{P}_{ab} \bigl(A = \pm 1 
			\ \text{and} \ B = \mp 1 \bigr) \ . 
	\end{equation}	
	Now recall that Gumbel's distribution has the following 
	joint cumulative distribution function 
	\begin{equation}
		F(x,y) = 1- e^{-x}-e^{-y}+e^{-x-y-xy}  
	\end{equation}
	(cf. Eq. 2.1 in \cite{gumbelBivariateExponentialDistributions1960}). 
	$p_{++}(a,b)$ is obtained from $F$ via a change of variables and 
	by summing over $\lambda_R$ (assuming $\xi > 0$): 
	\begin{equation}
		p_{++}(a,b) = \frac{1}{2} \,
			F\left(\frac{1/\xi- 1}{a+1},\frac{ 2 \left( 1/\xi- 1 \right)}{b-1} \right)
			 + \frac{1}{2} \, F\left(\frac{ 2 \left( 1/\xi- 1 \right)}{a-1} , 
			 \frac{1/\xi- 1}{b+1}
			 \right) \ . 
	\end{equation}
	Analogously, $p_{--}(a,b)$, $p_{+-}(a,b)$, and 
	$p_{-+}(a,b)$ are inferred from the common  
	identities for bivariate cumulative distribution functions. 

	The covariance is computed as in Eq. \eqref{eq:covAB}: 
	\begin{equation}
		\Cov_{a b} (A,B) = \Evalue(a,b) - 
			\bigl(2 p_{A+}(a)-1 \bigr) \bigl(2 p_{B+}(b)-1 \bigr) \ .
	\end{equation}	
	The respective quantities may be obtained from $p_{\pm \mp}(a,b)$ via 
	Eqs.  \eqref{eq:proof_jointprobs_1}, \eqref{eq:proof_jointprobs_2}, 
	and \eqref{eq:proof_jointprobs_4}. After a fair amount of 
	algebra, we find  
	\begin{multline}
		\Cov_{a b} (A,B)
		= 
		2 e^{\frac{(\xi-1) (\xi (2 a+b-1)+2)}{(a+1) (b-1) \xi^2}}
		+2 e^{\frac{(\xi-1) (\xi (a+2 b-1)+2)}{(a-1) (b+1) \xi^2}}
		-e^{\frac{(\xi-1) (2 a+b+1)}{(a+1) (b-1) \xi}}
		\\
		-e^{\frac{(\xi-1) (a+2 b+1)}{(a-1) (b+1) \xi}}
		-e^{\frac{2 (\xi-1) (a+b-2)}{(a-1) (b-1) \xi}}
		-e^{\frac{(\xi-1) (a+b+2)}{(a+1) (b+1) \xi}}
		\ . 
	\end{multline}
	Both for $\xi = 1$ and $\xi = 0$ the covariance vanishes, since 
	in that case $A$ and $B$ are constant. For $\xi \in (0,1)$ the covariance 
	is a nonzero function of $a$ and $b$. 
	
	In order to show that the BCHSH inequality 
	\eqref{eq:BCHSH} is nonetheless satisfied, we shall 
	rewrite $\Evalue(a,b)$. We first drop the obsolete dependence on 
	$\lambda_B$ and $\lambda_A$, so that the values of $A$ and $B$ 
	are ${A}(\lambda_R, \lambda_A)$ and ${B}(\lambda_R, \lambda_B)$. 
	Substituting $x=
	\lambda_A /(a+\lambda_R)$ and $y=\lambda_B / (b-\lambda_R)$ then 
	yields 
	\begin{multline}
		\Evalue(a,b) =
		\sum_{\lambda_R \in \lbrace \pm 1 \rbrace} 
		\int_{0}^\infty \d \lambda_A \int_{0}^\infty \d \lambda_B \,
		\rho_{ab} \left(\lambda_R, \lambda_A, \lambda_B \right) 
		{A}(\lambda_R, \lambda_A) \, {B}(\lambda_R, \lambda_B)
		\\
		= \frac{1}{2} \sum_{\lambda_R \in \lbrace \pm 1 \rbrace}
		\int_{0}^\infty \d x \int_{0}^\infty \d y \,
		f \left(x,y \right) 
		{A}\bigl( \lambda_R, x(a + \lambda_R) \bigr) \, 
		{B}\bigl(\lambda_R, y (b - \lambda_R) \bigr) \ . 
	\end{multline}
	This reduces the expression to the ``noncontextual'' case, 
	implying the assertion. 
\end{Example}

An example of a strongly local contextual Bell model violating 
the BCHSH inequality is given in the next section. 

In Ex. \ref{Ex:Gumbel} the correlation of $A$ and $B$ may be viewed as 
a consequence of the stochastic dependence of $\lambda_A$ and 
$\lambda_B$. In a reply \cite{gillKupczynskisContextualLocally2023} to 
Kupczynski \cite{kupczynskiEinsteinianNosignallingViolated2017,kupczynskiCanWeClose2017,
kupczynskiMoonThereIf2020,kupczynskiContextualityNonlocalityWhat2023}, 
Gill and Lambare argued that in such models 
``we are transferring the violation of local causality 
from the measurement outcomes to the instrument setting choices''. In other words, 
it is claimed that a stochastic dependence between the random detector 
parameters is necessarily a violation of ``locality''.

This objection is justified in the sense that in a strongly local contextual Bell model 
we may use ``nonlocal interactions'' in order to physically model such a stochastic dependence. 
In Ex. \ref{Ex:Gumbel}, for instance, the density in 
Eq. \eqref{eq:example_rhoab} may in principle originate from some 
``nonlocal communication'' between the detectors. 

Still, the general claim that a stochastic 
dependence between the random detector parameters always requires 
``nonlocal interactions'' lacks a clear-cut argument. In order 
to investigate the question mathematically, a rigorous, physically uncontroversial  
definition would be required. On the contrary, a single physically justified 
example of a strongly local contextual Bell model exhibiting such a stochastic 
dependence using only ``local interactions'' would invalidate the claim. 

The choice of the terminology ``strongly local'' in Def. 
\ref{Def:generalizedBM_locality} constitutes an implicit assertion that for general 
contextual Bell models this is the strongest notion of locality that 
does not exclude descriptions of Bell experiments relying only 
on ``local interactions''. In the case that a ``more refined'' mathematical 
description can be provided, it might, however, not be the only appropriate 
notion of locality. In Sec. \ref{sec:time} a physical discussion of such 
``more refined'' descriptions is given. 

\section{Violation of the BCHSH inequality in a strongly local contextual Bell model}
\label{sec:poc_example}

The purpose of this section is to provide a simple example 
of a strongly local contextual Bell model that violates the BCHSH 
inequality \eqref{eq:BCHSH}. As such, it provides a ``proof of concept'' 
that in a Bell experiment contextuality may lead to such a violation 
while the respective probabilistic model respects suitable conditions of locality. 

The contextual Bell model we consider is based on a family of discrete 
probability spaces. Recall that, by definition, a probability 
space is \emph{discrete}, if the sample space has at most countably many elements 
and the collection of random events is the respective power set. 
Specifically, we consider the case that 
$\Lambda_A$, $\Lambda_B$ and $\Lambda_R$ each contain only two elements. 

The following lemma shows that in any 
strongly local contextual Bell model of this type  
the families of probability measures  
are characterized by six functions and a constant. Note that for convenience we 
switch the order of $\Lambda_R$ and $\Lambda_A$ in the definition of $\Lambda$ in Eq.  
\eqref{eq:detector_decomposition_Lambda} and that we write 
$(i,j,k)$ instead of $(\lambda_A,\lambda_R, \lambda_B)$. 
\begin{Lemma}
	\label{Lem:2sample_ex}
	Let $\mathbb{B} = \lbrace 0, 1 \rbrace$, let $S$ be any nonempty set,  
	and let $\left( \mathbb{P}_{ab} \right)_{a,b \in S}$ 
	be a family of (discrete) probability measures 
	on $\mathbb{B}^3$ (together with its power set). For any 
	$i,j,k \in \mathbb{B}$, define 
	\begin{equation}
		q_{i j k} (a,b) = \mathbb{P}_{ab}\bigl( \lbrace (i,j,k) \rbrace \bigr) \ .   
	\end{equation}
	Furthermore, we require that $\sum_{i \in \mathbb{B}} q_{i j k} (a,b)$ and 
	$\sum_{k \in \mathbb{B}} q_{i j k} (a,b)$ are independent of the choice of 
	$a$ and $b$, respectively 
	(cf. property 
	\ref{itm:generalizedBM_stronglylocal}.\ref{itm:generalizedBM_stronglylocal1}
	in Def. \ref{Def:generalizedBM_locality}). 
	
	Then there exist a unique $c \in [0,1]$ and unique functions 
	$\gamma_l \colon S \times S \to [0,1]$, 
	$\alpha_l \colon S \to [0,1]$, as well as 
	$\beta_l \colon S \to [0,1]$
	with $l \in \mathbb{B}$
	such that for all $a,b \in S$: 
	\begin{subequations}
	\begin{align}
		q_{0 0 0} (a,b) &= \gamma_0(a,b) 
		\label{eq:2sample_ex_q000}
		\\
		q_{0 1 0} (a,b) &= \gamma_1(a,b) 
		\label{eq:2sample_ex_q010}
		\\
		q_{0 0 1} (a,b) &= -\gamma_0(a,b) + \alpha_0(a)
		\label{eq:2sample_ex_q001}
		\\
		q_{1 0 0} (a,b) &= -\gamma_0(a,b) + \beta_0(b) 		
		\label{eq:2sample_ex_q100}
		\\		q_{0 1 1} (a,b) &= -\gamma_1(a,b) + 1 - c - \alpha_1(a) 
		\label{eq:2sample_ex_q011}
		\\
		q_{1 1 0} (a,b) &= -\gamma_1(a,b) + 1 - c - \beta_1(b)
		\label{eq:2sample_ex_q110}
		\\
		q_{1 0 1} (a,b) &= \gamma_0(a,b) + c - \alpha_0(a) - \beta_0(b)
		\label{eq:2sample_ex_q101}
		\\
		q_{1 1 1} (a,b) &= \gamma_1(a,b) -1 + c + \alpha_1(a) + \beta_1(b) 
		\label{eq:2sample_ex_q111}
	\end{align}
	\end{subequations}
\end{Lemma}
\begin{Proof}
	The proof is constructive. First set  
	\begin{equation}
		\beta_{jk}(b) = q_{0 j k} (a,b) + q_{1 j k} (a,b)
		\quad \text{and} \quad
		\alpha_{ij}(a) = q_{i j 0} (a,b) + q_{i j 1} (a,b) \ . 
	\end{equation}
	Define $\gamma_0$ and $\gamma_1$ via Eqs. \eqref{eq:2sample_ex_q000}
	and \eqref{eq:2sample_ex_q010}. It follows that 
	\begin{subequations}
	\begin{align}
		q_{1j0} (a,b)= - \gamma_j (a,b)+ \beta_{j0} (b)\ , \\
		q_{0j1} (a,b) = - \gamma_j (a,b)+ \alpha_{0j}(a) \ . 
	\end{align}
	\end{subequations}
	Using an analogous derivation for $q_{1j1}$, we find the two conditions 
	\begin{equation}
		\alpha_{0j}(a) + \alpha_{1j}(a) = \beta_{j0}(b) + \beta_{j1}(b) \ .
		\label{eq:2sample_ex_proof_condition} 
	\end{equation}	
	As they must hold for all $a, b$, for each $j$  
	we hence obtain a constant $c_j$. By adding the conditions and recalling 
	normalization, we find $c_0=c$ and $c_1 = 1-c$ for 
	some $c \in [0,1]$. 
	
	We may define $\alpha_{0} = \alpha_{00}$, 
	$\alpha_{1} = \alpha_{11}$, $\beta_{0} = \beta_{00}$, and
	$\beta_{1} = \beta_{11}$. Hence $c$ determines the off-diagonal matrix 
	components via condition \eqref{eq:2sample_ex_proof_condition}. 
	Eqs. \eqref{eq:2sample_ex_q001} to \eqref{eq:2sample_ex_q110} 
	follow directly. Eqs. \eqref{eq:2sample_ex_q101} and 
	\eqref{eq:2sample_ex_q111} are obtained by recalling the 
	expressions for $q_{1j1}$. 
	
	Uniqueness is straightforward: first consider $q_{0j0}$ and 
	then $q_{100}$ as well as $q_{001}$. The latter fix $c$ via 
	Eq. \eqref{eq:2sample_ex_q101}. $\alpha_{1}$ and $\beta_{1}$ 
	must then also be unique due to Eqs. \eqref{eq:2sample_ex_q110} and 
	\eqref{eq:2sample_ex_q011}. 
\end{Proof}

In order to construct a contextual Bell model, (families of) detector random 
variables $A$ and $B$ need to be specified. As in Ex. \ref{Ex:correlatedP_cont}, 
we may choose the dependence on the setting parameters to be only implicit. 
For $i,j,k \in \mathbb{B}$ we take 
\begin{subequations}
	\label{eq:poc_example_AB}
\begin{align}
	A_{ij} &= 
		\begin{cases}
			+1 & , \, (i,j) = (0,0) \\
			-1 & , \, \text{else}
		\end{cases}
		\\
	B_{jk} &= 
		\begin{cases}
			+1 & , \, (j,k) = (0,0) \\ 
			-1 & , \, \text{else}
		\end{cases}
\end{align}
\end{subequations}
We hence obtain a strongly local contextual Bell model. 

The task is now to choose the constant and functions from Lem. 
\ref{Lem:2sample_ex} in such a manner that 
a violation of the BCHSH inequality \eqref{eq:BCHSH} is achieved.  

Via Eqs. \eqref{eq:poc_example_AB} and \eqref{eq:defpjoint}, 
we compute the following probability vector: 
\begin{multline}
	P(a,b) = \bigl( p_{++}(a,b), p_{-+}(a,b),p_{+-}(a,b),p_{--}(a,b) \bigr)
		\\
			= 
			\bigl( \gamma_0(a,b) , \beta_0(b)- \gamma_0(a,b) ,
				\alpha_0(a)-\gamma_0(a,b) ,1-\alpha_0(a)-\beta_0(b)+
				\gamma_0(a,b) \bigr) \ .
		\label{eq:poc_example_AB_Pvector}
\end{multline}
By Eq. \eqref{eq:proof_jointprobs_4} we have 
\begin{equation}
	\Evalue(a,b) = 1 - 2 \alpha_0(a) - 2 \beta_0(b) + 4 \gamma_0(a,b) \ . 
\end{equation}
The choice of $A$ and $B$ in Eq. \eqref{eq:poc_example_AB} hence implies that 
the values of $\Evalue(a,b)$ are determined via $\alpha_0$, $\beta_0$, and 
$\gamma_0$. 

In \cite{popescuQuantumNonlocalityAxiom1994} 
Popescu and Rohrlich provided example values for the probability vector $P(a,b)$ so that 
the maximal violation of the BCHSH inequality \eqref{eq:BCHSH} with a value of 
$4$ on the left hand side is achieved. 
We adapt their results to the conventions in this work, choosing  
$S= \mathbb{B}$, $a=b'=1$, and $b=a'=0$ for convenience. The respective values are 
\begin{subequations}
\begin{gather}
	P(0,0) = P(0,1) = P(1,0) =\left( 1/2, 0, 0, 1/2 \right) \ , 
	\\
	P(1,1) = \left(0, 1/2, 1/2, 0 \right) \ . 
	\label{eq:PopescuRohrlich}
\end{gather}
\end{subequations}
Comparing with Eq. \eqref{eq:poc_example_AB_Pvector}, the following choices provide the 
required values: 
\begin{subequations}
\begin{gather}
\alpha_0(0) = \beta_0(0)= \alpha_0(1) = \beta_0(1)= 1/2 \\
\gamma_0(0,0) = \gamma_0(1,0) = \gamma_0(0,1) = 1/2 \\
\gamma_0(1,1)= 0
\end{gather}
\end{subequations}
Furthermore, for $q_{101}(1,1)$ to stay positive, we require that $c=1$. Similarly, 
positivity of $q_{011}$ and $q_{110}$ forces the functions $\gamma_1$, 
$\alpha_1$, and $\beta_1$ to vanish. 

We hence find the following 
probabilities for the elementary events: 
\begin{subequations}
\begin{gather}
	q_{001}(1,1)=q_{100}(1,1)=1/2 \ , \\
	q_{000}(a,b)=q_{101}(a,b)=1/2 \quad \text{for} \ (a,b) \in 
	\lbrace (0,0),(1,0),(1,0) \rbrace \ ,
\end{gather} 
\end{subequations}
and all remaining $q_{ijk}(a,b)=0$. It is straightforward to show that for any 
possible choice of settings the random detector parameters $i$ and $k$ are 
stochastically dependent. 

In this case, we indeed have 
\begin{equation}
	\Evalue(1,0)+\Evalue(0,1)+\Evalue(0,0)-\Evalue(1,1) = 4 \ .
\end{equation}
The right hand side is above the 
Tsirel'son bound \cite{cirelsonQuantumGeneralizationsBells1980,
peresQuantumTheoryConcepts2002} 
of $2 \sqrt{2} \approx 2.828$ (cf. Cor. \ref{Cor:BCHSHcontra}). 

Since this contextual Bell model is strongly local 
and also violates the BCHSH inequality, there cannot be 
any equivalent ``noncontextual'' description that retains the 
property of strong locality (cf. Cor. \ref{Cor:stronglocalityBCHSH}). 

Of course, the example here was derived from purely mathematical 
considerations. 

In this regard, it is noteworthy that $j=0$ with 
probability $1$ for any setting, so we may construct an ``equivalent'' contextual 
Bell model on $\mathbb{B} \times \lbrace 0 \rbrace \times \mathbb{B}$ with 
sample probabilities given by Eq. \eqref{eq:PopescuRohrlich}. 
Recalling Fig. \ref{fig:Bellexperiment}, we 
may think of a Bell experiment for which the detectors have a single switch 
with settings $0$ (``off'') and $1$ (``on'') 
as well as red and green output lights. Regardless 
of the setting, for each trial there is always a $50 \%$ chance that any detector 
shows a red or green light. If all switches are in the off-position or only one of 
them is in the on-position, the lights always have the same colors. Yet once 
both switches are in the on-position, the lights always have opposing colors. 

What is shown here is that this \emph{Popescu-Rohrlich (thought) experiment}  
\cite{popescuQuantumNonlocalityAxiom1994} may be described by a strongly local 
contextual Bell model. Yet as in the usual model for the throw of a 
fair six-sided die, no physical justification is provided for how those probabilities 
arise (cf. Secs. \ref{sec:discussion}, \ref{sec:time}, as well as the discussion at 
the end of Sec. \ref{sec:generalizedBell}). 

Accordingly, a physically motivated example of a strongly local 
contextual Bell model that is based on ``locally mediated interactions'' and that 
also violates the BCHSH inequality remains to be found. 

\begin{Remark}
	\label{Rem:Arroyo_model}
	An example of a contextual Bell model violating Bell's inequality \eqref{eq:Bellineq} 
	has previously been given by Arroyo 
	\cite{arroyoFamilyDeterministicModels2025}. However, the proposed model is 
	not strongly local, since the distribution 
	$\mathbb{P}_{ab}\circ \pr_R^{-1}$ depends on the detector settings. 
	Similarly, the examples given in \cite{feldmannNewLoopholeEinsteinPodolskyRosen1995} 
	are not strongly local. Another example by Feldmann 
	\cite{feldmannClassicalCounterexampleBells2010} 
	is weakly local, yet does not appear to be strongly local. 
\end{Remark}

\section{On temporal evolution in Bell experiments}
\label{sec:time}

In Sec. \ref{sec:generalizedBell} we saw how physical considerations 
motivate a generalization of the concept of a Bell model. 
For this purpose, we introduced the notion of a contextual Bell model 
(Def. \ref{Def:generalizedBM}) along with corresponding definitions of 
strong and weak locality (Def. \ref{Def:generalizedBM_locality}). 
In Sec. \ref{sec:poc_example} we 
found that such contextual Bell models can violate the BCHSH inequality, even 
if they are strongly local. 
The purpose of this section is to elaborate on the question of how a ``local'' 
theory of nature based on mathematical probability theory -- such as 
geometric quantum theory -- can predict a violation 
of the BCHSH inequality in specific instances. Specifically, we argue 
that time evolution takes a central role 
in how the required correlation between the (families of) detector random variables 
may arise. 

There are two reasons why, so far, we have mostly ignored the role of time 
evolution in Bell experiments. 

First, in Sec. \ref{sec:discussion} we argued that 
the four joint 
probabilities $p_{\pm \mp}(a,b)$ 
suffice for a basic 
phenomenological description of a given Bell experiment 
with settings $a$ and $b$. This argument carries over to Bell experiments, which are described by contextual Bell models. 
Stated plainly, it is these joint probabilities that are ultimately relevant for testing 
a potential violation of the BCHSH inequality in a given experiment---not the underlying time evolution. 

Second, in the archetypal Bell experiment discussed in Sec. \ref{sec:review}, 
the contemporary, conventional view asserts  
that the temporal evolution of the system can be 
ignored up to the point that the particles are 
``measured'' \cite{bohmDiscussionExperimentalProof1957}: 
we are interested in a $2$-body quantum system in which the Hamiltonian 
commutes with the spin operators, so that the spin components are conserved 
throughout time (e.g. due to the absence of magnetic fields away from the detectors). 
Once the two ``classical measurement apparatus'' make a ``measurement'', the 
state is projected. Then, over a sufficiently large number of trials 
the prediction for the averages \eqref{eq:naiveQM} should hold. 

With regards to the first point, time evolution is nonetheless of interest once one   
aims to theoretically explain a potential violation of the BCHSH inequality in 
a given experiment. 
A corresponding contextual Bell model needs to be justified on the basis of 
physical theory; 
it is to be derived from the dynamical evolution and the initial conditions of 
the system as a whole (viewed as an ensemble): 
the particles, the detectors and anything else of relevance 
to their dynamics. 

With regards to the second point, it is the quantum-mechanical 
formalism itself that discounts the argument as naive. Indeed, Bell already noted this 
in Sec. 3 of his 1971 article \cite{bellIntroductionHiddenvariableQuestion1971,
bellSpeakableUnspeakableQuantum2004}. Therein, he 
emphasized that the measurement apparatus is to be included in the formulation of 
the dynamical problem, 
at least for the purpose of investigating foundational questions: 
\begin{quote}
	The result of a ``spin measurement'', for example, depends in a very complicated 
	way on the initial position [...] of the particle and on the strength 
	and geometry of the magnetic field. Thus, the result of the measurement 
	does not actually tell us about some property previously possessed by the 
	system, but about something which has come into being in the combination of 
	system and apparatus.   
\end{quote}
That is, if we accept the dynamical equations of quantum theory, we must reject 
modeling measurement 
by such a projection as overly simplistic. The averages in 
Eqs. \eqref{eq:naiveQM} and \eqref{eq:GHSZ_exp} 
are to be viewed as rule-of-thumb approximations to actual quantum-mechanical predictions 
at best, irrespective of whether they violate Bell inequalities or not. 
A corresponding, general argument against this naive view of measurement via projection 
has been put forward by Ballentine \cite{ballentineStatisticalInterpretationQuantum1970,
ballentineLimitationsProjectionPostulate1990}. 
Indeed, much of Bell's research was motivated by the lack of ``physical precision'' 
of such rule-of-thumb predictions 
\cite{bellMeasurement1990,bellSpeakableUnspeakableQuantum2004}. 

In summary, a theoretical justification for an empirically confirmed violation of the 
BCHSH inequality in a given experiment requires a concrete dynamical model for the system 
as a whole. This includes, in particular, a dynamical model for the detectors.

It is, however, beyond the scope of this article to provide such a 
justification for specific cases, for which the 
empirical violation of the BCHSH inequality 
has been shown convincingly. 

Rather, we shall address the following 
fundamental question, continuing the discussion at 
the end of Sec. \ref{sec:generalizedBell}: Can a 
physically justified, strongly local contextual Bell model violate the BCHSH 
inequality without the introduction of ``nonlocal interactions''? 

For this purpose, we address the general question of how 
the system as a whole is to be modeled dynamically 
without resorting to rule-of-thumb arguments.

From part I 
\cite{reddigerApplicabilityKolmogorovsTheory2025,
reddigerAddendumApplicabilityKolmogorovs2026} of this series, 
we recall that quantum mechanics and (non-relativistic) geometric quantum theory 
are mostly distinguished by the underlying probability theory they employ; they 
generally agree on how a particular system is to be modeled dynamically, irrespective of 
some subtle questions of regularity. That is, at least in principle and following 
Bell's above remark, the two theories ought to agree on the 
dynamical model for the measurement process---even for the most complicated measurement setups. 
This is, of course, assuming that we abstain from a simplified modeling of the 
measurement process via projection in either theory. 

This assessment is, however, not taking account of the need to distinguish between 
non-relativistic and relativistic quantum theory. Our task is therefore 
to refine this view in the two different contexts. 

With regards to non-relativistic quantum dynamics, 
the following is known \cite{galindoPropagacionInstantaneaSistemas1968,
hegerfeldtCausalityParticleLocalization1998,
madridLocalizationNonRelativisticParticles2007,
beckLocalQuantumMeasurement2021}: if the Hamiltonian is 
bounded from below and the initial (position) wave function is  
supported in a bounded region, then for future times it either remains 
supported in that region or it will spread ``infinitely fast''. 
At least heuristically, it may therefore be expected that in non-relativistic quantum 
theory ``nonlocal'' behavior is the norm, not the exception. 
That is, contextual Bell models derived from non-relativistic dynamics need neither 
respect strong nor weak locality. And from any such violation in the model, no conclusion 
can be drawn on the physical violation of ``locality'' in the Bell experiment the model 
aims to describe. 

We are hence forced to investigate the question of ``locality'' in the relativistic setting. 
And in relativistic quantum theory, dynamics is a more delicate topic. 

While the axiomatic foundations of quantum mechanics 
\cite{vonneumannMathematicalFoundationsQuantum1955,morettiFundamentalMathematicalStructures2019}
and non-relativistic quantum field theory \cite{araiAnalysisFockSpaces2018} 
have been established, the debate on the foundations of relativistic quantum theory 
remains unsettled  \cite{fraserQuantumFieldTheory2009,wallaceTakingParticlePhysics2011,
fraserPerturbativeExpansionsFoundations2024,reddigerProbabilisticFoundationRelativistic2024,
fraserEnactingRigourGlobalRebuilding2026}. 
Only in the 1950s did the mathematical physics 
community begin to address the axiomatization of 
\mbox{(special-)}relativistic quantum theory  
(see e.g. 
\cite{friedrichsMathematicalAspectsQuantum1953,
	friedrichsMathematicalAspectsQuantum1951, 
	friedrichsMathematicalAspectsQuantum1952,
	friedrichsMathematicalAspectsQuantum1952a,
	friedrichsMathematicalAspectsQuantum1953a,
	wightmanQuantumFieldTheory1956,wightmanFieldsOperatorvaluedDistributions1964,
	cookMathematicsSecondQuantization1953,
	lehmannZurFormulierungQuantisierter1955,schmidtQuantentheorieFelderAls1956,
	jostIntegralDarstellungKausalerKommutatoren1957,arakiAsymptoticBehaviourWightman1962}). 
Arguably, the most developed approach in contemporary physics is algebraic quantum field theory 
(see e.g. \cite{haagLocalQuantumPhysics1996,rejznerPerturbativeAlgebraicQuantum2016,
brunettiAdvancesAlgebraicQuantum2015,AlgebraicQuantumField2025}). 
The theory is based on the Haag-Kastler 
axioms \cite{haagAlgebraicApproachQuantum1964} and has also been generalized to 
curved spacetime (see e.g. \cite{brunettiGenerallyCovariantLocality2003,
sandersAspectsLocallyCovariant2008,barQuantumFieldTheory2009,
hollandsAxiomaticQuantumField2009,
hollandsQuantumFieldsCurved2015,brunettiAdvancesAlgebraicQuantum2015,fewsterQuantumFieldsLocal2020,
fischerContinuumLimitAnalysis2026}). 

With regards to the applicability of mathematical probability theory to 
quantum phenomena, the question of relativistic 
dynamics is subtle for the following reason: commonly employed axiomatic approaches 
to relativistic quantum theory rely heavily on a generalization of  
concepts in quantum probability theory. In algebraic quantum field theory, 
for instance, the philosophical 
principles of ``locality'' and ``causality'' are implemented via commutation 
relations 
(cf. \cite{haagAlgebraicApproachQuantum1964} and 
p. 108 in \cite{haagLocalQuantumPhysics1996}). On the contrary, if 
we base relativistic quantum theory on Kolmogorov's axioms, such an 
approach lacks justification: here 
``observables'' are real- or vector-valued random variables on a suitable 
probability space, not (densely defined) operators on a Hilbert space 
or ``operator-valued distributions'' 
(see e.g. Sec. II.1.2 in \cite{haagAlgebraicApproachQuantum1964} or 
Sec. 8.2 in \cite{araiAnalysisFockSpaces2018}). Contrary to 
non-relativistic quantum theory, a change of the probability theory in 
the relativistic theory 
has such structural consequences that one cannot simply carry over the dynamics 
from approaches that are based on generalizations of quantum-mechanical 
concepts. An illustration thereof may be found in 
\cite{reddigerProbabilisticFoundationRelativistic2024}. Therein,   
the special case of a one-body theory in curved spacetime is considered. 
Further research is needed before ab initio models of 
simple Bell experiments are feasible. 

Despite the contemporary state of progress in the probabilistic foundations 
of relativistic quantum theory, we may still infer some statements on 
contextual Bell models in the relativistic setting. 

Since (contextual) weak locality is consistent with observations, 
any plausible relativistic quantum theory -- whatever it may be -- 
is expected to make predictions that comply with it. 
Similarly, if Bell inequalities 
are indeed empirically violated for certain quantum systems 
-- a basic assumption of this work -- then such a violation 
ought to be a genuine dynamical prediction. 

It was shown at the end of Sec. \ref{sec:generalizedBell} 
that such a violation requires a correlation of the (families of) 
detector random variables $A$ and $B$ as a necessary condition. 
We hence have to ask the question of the underlying mechanism 
in the context of relativistic 
quantum theory. Though this may only receive a 
speculative answer at this point, there are some nonrigorous, qualitative 
arguments one can make. 

Let us suppose that initially -- i.e. on a suitable nowhere timelike hypersurface -- 
the random variables associated to
the particle going to detector $A$ are stochastically independent from the random 
variables associated to the particle going to detector $B$. 
This is already an idealizing assumption, but we shall make it nonetheless. 

Now, from non-relativistic quantum theory it is known, 
that in the presence of interactions 
a many-body system initially described by a tensor product of 
one-body wave functions will usually evolve to a many-body wave function 
for which this factorization no longer holds. 
In other words, interactions tend to introduce a stochastic 
dependence between the positions of the different particles and hence other 
random variables associated to the particles. Physically, this may be understood 
as a consequence of the Bohm force/Bohm potential, see e.g. Sec. 
7.2.1 in \cite{hollandQuantumTheoryMotion1993}. Plausibly, a 
physically suitable relativistic quantum theory does not change this 
qualitative behavior. That is, in the presence of interactions 
time evolution tends to induce a stochastic dependence both between the random 
variables associated to
the particle going to detector $A$ and the random 
variables associated to the particle going to detector $B$. And because 
the detectors themselves consist of particles subject to quantum dynamics and 
interacting with the incoming particle, 
here, too, we expect a stochastic dependence between the random 
variables associated to the incoming particle at the time of 
detection and the respective random detector 
parameter. Though this does not necessarily imply a correlation 
between $A$ and $B$ for certain detector settings, dynamical evolution of the system 
as a whole can make this correlation at least plausible. 

The qualitative, physical argument is therefore as follows: 
due to quantum dynamics and interactions, 
stochastic dependence between random variables 
associated to the particles as well as the detectors is the norm, not the exception. 
And under certain conditions this may lead to correlations that imply 
a violation of Bell inequalities. 

\begin{Remark}
	\label{Rem:reldynamics}
\begin{enumerate}[1)]
	\item 	\label{itm:Rem:reldynamics1}
			There is a loose, mathematical meaning one can give to 
			the requirement of ``locally mediated interactions'':  
			the time evolution of the system as a whole is to be determined 
			by equations, which are formulated 
			in terms of local operators only. For instance, 
			partial differential evolution equations would be appropriate, 
			because differential operators are local in the sense that the 
			value of a derivative of any order at a point only depends on 
			the values of the map in a neighborhood of that point. 
			Contrarily, integro-differential equations ought not to be 
			accepted as fundamental dynamical laws, since integral operators 
			are usually nonlocal in that sense. 
			
			One may speculate that for detectors, which 
			are separated sufficiently far, weak locality -- and perhaps 
			even strong locality -- may be a consequence of 
			this mathematical notion of locality in the (relativistic) 
			evolution equations. 
			Nonrigorous attempts to draw a connection between 
			``locally mediated interactions'' and Bell inequalities in a 
			relativistic setting 
			were made in \cite{bransBellsTheoremDoes1988,
			zychBellsTheoremTemporal2019}. 
	
			It is worth pointing out that the above dynamical notion of locality 
			is very broad 
			and would, for instance, include (relativistic generalizations of) 
			de Broglie-Bohm theory (see e.g. 
			\cite{hollandQuantumTheoryMotion1993,bohmUndividedUniverseOntological1993,
			durrHypersurfaceBohmDiracModels1999}).
	\item 	\label{itm:Rem:reldynamics2}
			Hess and Philipp 
			\cite{hessEinsteinseparabilityTimeRelated2001,
			hessPossibleLoopholeTheorem2001,hessExclusionTimeTheorem2002} 
			have also argued that (relativistic) time evolution may give 
			rise to contextual Bell models violating Bell inequalities 
			(see also \cite{larssonBellsInequalityCoincidencetime2004}). 
			However, their argument differs from the one provided here. 
	\item 	\label{itm:Rem:reldynamics3}
			Rem. \ref{Rem:reldynamics}.\ref{itm:Rem:reldynamics1} 
			is subject to one caveat: 
			If we understand a contextual Bell model as resulting from 
			a solution of a corresponding dynamical problem within quantum theory, 
			then the setting parameters 
			$a$ and $b$ may not merely refer to the 
			respective setting of the detector at the time of ``measurement''. 
			Rather, they must encode the settings for a certain duration, 
			as dictated by the dynamical model.  
			Usually it will, at least, include the time span from which 
			the particles are ejected from the ``particle gun'' up until 
			both detectors have provided a reading. 

			Hence, we should generally not view an element $a \in S$ as the 
			setting at the time of detection, but as a curve 
			$a \colon t \mapsto a(t)$ in some set $S'$, 
			so that $a(0) \in S'$ denotes the setting at the time of ejection 
			and $S$ is a suitable collection of curves in $S'$. 
			
			Indeed, this point is of experimental importance, 
			for several Bell test experiments 
			vary the setting (pseudo-)randomly throughout the run and then 
			the results are assigned to the same ensemble, whenever said setting 
			is the same at the time of detection. Unless care is taken  
			that this random change of the setting can 
			``average out'', this practice is problematic 
			due to a violation of the ``similarly prepared systems'' assumption 
			(see Sec. \ref{sec:statistics} below). 
			At least in principle, the detector settings may physically influence 
			the measurement result before detection.
\end{enumerate}
\end{Remark}

\section{The connection to statistical data}
\label{sec:statistics}

In order to empirically determine a potential violation of the BCHSH inequality, 
the experimentalist 
needs to conduct multiple trials for the four chosen setting combinations. 
The experimentalist cannot measure the 
expectation value $ \Evalue(a, b )$ directly---it is only indirectly accessible through 
statistical reasoning. Drawing this connection is a standard exercise in 
mathematical statistics \cite{schervishTheoryStatistics1997}. Still, we  
shall carry it out here in some detail, for it might not be obvious 
to the reader how to do so and the precise connection of the theory to experiment 
is of great importance for drawing conclusions from statistical data. The main 
tool is the central limit theorem for independent and identically 
distributed random variables. We also give an example showing 
that statistical Bell inequality violations can occur even in strongly local 
``noncontextual'' Bell models, if an insufficient amount of data is collected. 

We carry out our analysis for a general contextual Bell model 
(Def. \ref{Def:generalizedBM}) without 
imposing any requirements of locality (Def. \ref{Def:generalizedBM_locality}). 
For any $a,b \in S$, we further define 
the random variable $X({a}, {b})$ via Eq. \eqref{eq:defX_general}. Its expectation 
value is given via Eq. \eqref{eq:Eabsplit_full} or equivalently 
\begin{equation}
	\Evalue(a,b) = \int_{\Lambda} \d \mathbb{P}_{ab} \, 
						 X(a,b)  \ .
	\label{eq:Eab_contextual}
\end{equation}

Again, for a single trial with settings $a$, $b$ and label 
$\lambda$, the experimentalist records the values $A(a,b)(\lambda)$ and 
$B(a,b)(\lambda)$. These are multiplied to determine $X({a}, {b})(\lambda)$. 

The first step is to model this process for multiple trials with different 
choices of settings. In general, this model depends on the exact order 
of which different combinations of settings are chosen and how many trials for each 
combination of settings are carried out. However, if the experiments 
are carried out in a manner that the ``similarly 
prepared systems'' assumption may be assumed to hold, then the order is 
ultimately irrelevant for the statistical analysis. Moreover, we may make the 
simplifying assumption that for 
each combination of settings the number of trials is the same. 

Hence we assume that the settings are chosen in the order 
$(a,b)$, $(a',b')$, $(a',b)$, and $(a,b')$ and that a total number of 
$n \in \N$ such rounds of four trials 
are carried out.

Rather than setting up a probability space for these $4n$ trials, however, we follow the  
custom in mathematical statistics by considering 
a countably infinite number of trials. We then only compute the relevant 
quantities for the first $4n$ trials. This 
eases the study of the limit for large $n$. 
Accordingly, we consider the 
sample space 
\begin{equation}
	\Lambda^\infty = \bigtimes_{i=1}^\infty \Lambda 
\end{equation}
along with the infinite product $\sigma$-algebra 
\begin{equation}
	\mathcal{A}^\infty = \bigotimes_{i=1}^\infty \mathcal{A}  \ . 
\end{equation}
The space $\Lambda^\infty$ may be understood as the space of infinite sequences 
$i \mapsto \lambda_i$ in $\Lambda$. To define the  
probability measure on $(\Lambda^\infty,\mathcal{A}^\infty)$, 
we first consider the product measure 
\begin{equation}
	\label{eq:statistics_Pround}
	\mathbb{P}_{a b a' b'} = 
		\mathbb{P}_{a b} \otimes \mathbb{P}_{a' b'} \otimes 
		\mathbb{P}_{a' b} \otimes \mathbb{P}_{a b'}
\end{equation}
on $\left(\Lambda^4, \mathcal{A}^4\right)$ for a single round of trials. We may 
then formally construct the (infinite) product measure  
\begin{equation}
	\label{eq:statistics_Pinfty}
	\mathbb{P}_{a b a' b'}^\infty = \bigotimes_{n=1}^\infty \mathbb{P}_{a b a' b'} \ . 
\end{equation}
The following lemma assures that the above construction is well-defined. 

\begin{Lemma}
	\label{Lem:contextualBellstatistics}
	Consider a contextual Bell model as in Def. \ref{Def:generalizedBM} and define 
	$\Lambda^\infty$, $\mathcal{A}^\infty$, as well as 
	$\mathbb{P}_{a b a' b'}$ as above for any $a, b, a', b' \in S$. 
	
	Then the probability measure  
	$\mathbb{P}_{a b a' b'}^\infty$ on $\left( \Lambda^\infty, 
	\mathcal{A}^\infty\right)$ from Eq. \eqref{eq:statistics_Pround} 
	is uniquely determined by the following condition: 
	for any $n \in \N$ and any sequence of events  
	$U_1, \dots, U_n$ in $\mathcal{A}^4$ it holds that 
	\begin{equation}
		\mathbb{P}_{a b a' b'}^\infty \left( U_1 \times \dots \times U_n 
					\times \bigtimes_{i=n+1}^\infty \Lambda \right) 
			= \prod_{i=1}^n \mathbb{P}_{a b a' b'} \left( U_i \right) \ . 
	\end{equation}
	
	Furthermore, for each $i \in \N$ the coordinate maps 
	\begin{equation}
			\Lambda^\infty \to \Lambda \ , \quad \boldsymbol{\lambda} 
			= \left(\lambda_1, \dots, \lambda_i, \dots \right)
			\mapsto 
			\lambda_i 
	\end{equation}
	are stochastically independent under $\mathbb{P}_{a b a' b'}^\infty$. 
\end{Lemma}
\begin{Proof}
	This is a consequence of Ionescu-Tulcea's theorem, see  Cor. 14.36 in 
	\cite{klenkeProbabilityTheoryComprehensive2020}. 
\end{Proof}

Given the probability space $\left( \Lambda^\infty, 
	\mathcal{A}^\infty,\mathbb{P}_{a b a' b'}^\infty \right)$, we may define  
the random variable $Y_i ({a}, {b},{a}', {b}')$ 
for round number $i \in \N$ via 
\begin{multline}
	Y_i ({a}, {b},{a}', {b}')(\boldsymbol{\lambda} ) 
		= X({a}, {b})( \lambda_{4i-3}) 
			+ X({a}', {b}')( \lambda_{4i-2})
			\\ 
			+ X({a}', {b})( \lambda_{4i-1})
			- X({a}, {b}')( \lambda_{4i} )  
		\label{eq:defYi_statistics}
\end{multline}
with $\boldsymbol{\lambda} \in \Lambda^\infty$. The random variables 
$Y_1, \dots, Y_n , \dots$ are 
mutually independent by Lem. \ref{Lem:contextualBellstatistics}. They are identically 
distributed by construction. 

For testing the BCHSH inequality \eqref{eq:BCHSH} statistically, 
the quantity of primary interest is 
the arithmetic mean of the $Y$s after $n$ rounds: 
\begin{equation}
	\bar{Y}_n({a}, {b},{a}', {b}') =  \frac{1}{n} \sum_{i=1}^n Y_i({a}, {b},{a}', {b}')
	\ . 
\end{equation} 
In general, its absolute value is bounded by $4$. Accordingly, 
for given $\boldsymbol{\lambda}$ the measured 
quantity is 
\begin{equation}
	\bar{Y}_n({a}, {b},{a}', {b}')(\boldsymbol{\lambda}) \ . 
	\label{eq:BCHSH_statistics_measuredY}
\end{equation}
We say that the BCHSH inequality 
is \emph{statistically violated} in a given Bell experiment, if 
the expression \eqref{eq:BCHSH_statistics_measuredY} is strictly above $2$. 
Otherwise, it is \emph{statistically valid}. This terminology suffices for our 
purposes here; more commonly one would tie its usage to a given statistical 
confidence level. 

The main tool for relating the expression \eqref{eq:BCHSH_statistics_measuredY} 
to the left hand side of the BCHSH inequality \eqref{eq:BCHSH} 
is the central limit theorem 
(see also \cite{feldmannNewLoopholeEinsteinPodolskyRosen1995}). 

\begin{Proposition}[Central limit theorem, adapted]
	\label{Prop:contextualBellstatistics_clt}
	Under the conditions of Lem. \ref{Lem:contextualBellstatistics}, let 
	$X$, $Y$, and $\bar{Y}$ be given as above. Furthermore, for 
	$a,b \in S$ denote by $\Var_{ab}(Z)$ the variance of any random variable $Z$
	with respect to $\mathbb{P}_{ab}$. Define $\Evalue(a,b)$ via Eq. 
	\eqref{eq:Eab_contextual} and set 
	\begin{equation}
		\sigma_{ab} = \sqrt{\Var_{ab}\left( X(a,b) \right)} \ . 
	\end{equation}
	Denote by $\mathcal{N}_{0,1}$ the standard normal distribution on 
	$\left(\R,\mathcal{B}(\R)\right)$. 
	
	Then the distribution of the (standardized) random variable 
	\begin{multline}
		\bar{Y}^*_n({a}, {b},{a}', {b}') = 
		\sqrt{\frac{n}{\sigma^2_{ab} + \sigma^2_{a'b'} + 
			\sigma^2_{a'b} + \sigma^2_{ab'}}} 
			\\ 
		\Bigl( \bar{Y}_n({a}, {b},{a}', {b}') - 
			\left(\Evalue(a,b) + \Evalue(a',b') + \Evalue(a',b) - \Evalue(a,b')\right)
		\Bigr)
			\label{eq:Yi_statistics_standardized}
	\end{multline}
	satisfies the following for any Borel set $I \in \mathcal{B}(\R)$:  
	\begin{equation}
		\lim_{n \to \infty} \mathbb{P}_{a b a' b'}^\infty 
			\Bigl( \bar{Y}^*_n({a}, {b},{a}', {b}') \in I \Bigr)
			= \mathcal{N}_{0,1} (I) \ . 
	\end{equation}
\end{Proposition}
\begin{Proof}
	We shall apply the central limit theorem 
	(see e.g. Thm. 15.38 in \cite{klenkeProbabilityTheoryComprehensive2020}), 
	so we merely need to compute the expectation value and the 
	variance of any $Y_i ({a}, {b},{a}', {b}')$ with respect to the measure 
	$ \mathbb{P}_{a b a' b'}^\infty $. 
	
	The expectation value is given by 
	\begin{equation}
		\mathbb{E}_{a b a' b'}^\infty \left( Y_i ({a}, {b},{a}', {b}')\right)
			= \Evalue(a,b) + \Evalue(a',b') + \Evalue(a',b) - \Evalue(a,b') \ . 
			\label{eq:clt_E}
	\end{equation}
	It exists due to boundedness of $X$. 
	
	For the variance, we first recall that, due to Lem. 
	\ref{Lem:contextualBellstatistics} and Eq. \eqref{eq:statistics_Pround}, 
	the coordinate maps 
	are stochastically independent. Hence, the individual summands on 
	the right hand side of Eq. \eqref{eq:defYi_statistics} are stochastically 
	independent and a simplified formula for the variance holds: 
	\begin{align}
		\Var_{a b a' b'}^\infty \left(  Y_i  ({a}, {b},{a}', {b}')\right)
			&= \Var_{ab}\left( X(a,b) \right) + \Var_{a'b'}\left( X(a',b') \right) 
			\notag
			\\
			& \quad  \quad 
			+ \Var_{a'b}\left( X(a',b) \right) + \Var_{ab'}\left( X(a,b') \right)
			 \\
			&=\sigma^2_{ab} + \sigma^2_{a'b'} + 
			\sigma^2_{a'b} + \sigma^2_{ab'} 
			 \ . 
			 \label{eq:clt_Var}
	\end{align}
	Boundedness of $X$ again guarantees that the latter quantities exist. 
\end{Proof}

It follows from Prop. \ref{Prop:contextualBellstatistics_clt} that for large 
$n$ we may approximate the distribution of the mean 
$\bar{Y}_n({a}, {b},{a}', {b}')$ by a normal distribution with 
expectation value \eqref{eq:clt_E} and standard deviation 
\begin{equation}
 	\sqrt{\left(\sigma^2_{ab} + \sigma^2_{a'b'} + 
			\sigma^2_{a'b} + \sigma^2_{ab'}\right)/n } \ . 
\end{equation} 
For ``sufficiently large'' $n$, this standard deviation is small and 
it is thus very likely that the measured quantity 
\eqref{eq:BCHSH_statistics_measuredY} is close to the 
expectation value. 

Since one does generally not know the distribution of the $X$s, 
the meaning of the phrase ``sufficiently large'' is difficult to 
quantify. Given the second and third moments 
of that distribution, the 
Berry-Esseen theorem provides 
an estimate of the speed of convergence 
(see e.g. Thm. 15.52 in \cite{klenkeProbabilityTheoryComprehensive2020}). 
Without that knowledge, the Chernoff-Hoeffding bound provides a gross estimate for  
the minimum number of rounds. 
\begin{Lemma}[Chernoff-Hoeffding bound, adapted]
	\label{Lem:Hoeffding}
	Under the conditions of Lem. \ref{Lem:contextualBellstatistics}, let 
	$\bar{Y}$ be given as above and denote by $\mu(a,b,a',b')$ the right hand side of 
	Eq. \eqref{eq:clt_E}. Then for 
	all $a,b,a',b' \in S$, $\varepsilon > 0$, and $n \in \N$ we have 
	\begin{equation}
		\mathbb{P}_{a b a' b'}^\infty \Bigl( 
		\abs{ \bar{Y}_n({a}, {b},{a}', {b}') - 
		\mu(a,b,a',b')} \geq \varepsilon
		\Bigr) \leq 2 e^{- n \varepsilon^2 / 32}
		\ . 
		\label{eq:Hoeffding}
	\end{equation}
\end{Lemma}
\begin{Proof}
 	See e.g. Thm. 4.12 in \cite{michaelmitzenmacherProbabilityComputing2012}, 
 	noting that $\abs{Y} \leq 4$. 
\end{Proof}

Denote by $x \mapsto \lceil x \rceil$ the ceiling function. 
Given a maximum error $\varepsilon$ and an upper bound on the (error) 
probability $\gamma$ on the left hand side of Eq. \eqref{eq:Hoeffding}, 
we hence find that 
\begin{equation}
	n_0 = \left\lceil  \frac{32}{\varepsilon^2} \, \ln (2/ \gamma) \right\rceil  
\end{equation}
provides a corresponding estimate for the least number of rounds. Note that 
this is not a statement on the validity of the aforementioned approximation via Prop. 
\ref{Prop:contextualBellstatistics_clt}. 

In practice, the Chernoff-Hoeffding bound yields rather large values of 
$n_0$. So instead one assumes that the approximation via 
Prop. \ref{Prop:contextualBellstatistics_clt} is justified for the given number of 
rounds $n$. This allows one to compute, for instance, confidence intervals 
for the quantity \eqref{eq:clt_E} for a 
normal distribution with unknown mean and unknown variance. The reader is referred, 
for instance, to Sec. 1.1.2 and Ex. 5.94 in \cite{schervishTheoryStatistics1997} 
or to any other textbook on mathematical statistics. 

On the basis of either Lem. \ref{Lem:Hoeffding} or Prop. 
\ref{Prop:contextualBellstatistics_clt}, an empirical (non)violation of the 
BCHSH inequality 
\eqref{eq:BCHSH} may be inferred. While the former method is rigorous and 
the latter is more practical, either 
one of the two methods requires a choice 
on the desired degree of statistical confidence. 
It is the nature of statistics that there is no absolute certainty.

From an experimental perspective, it is noteworthy that the ``similarly 
prepared systems'' condition is mathematically implemented by Eqs. 
\eqref{eq:defYi_statistics}, \eqref{eq:statistics_Pround}, and 
\eqref{eq:statistics_Pinfty}, so that the outcomes for each trial are mutually 
independent and identically distributed. 
While there do exist variants of the central limit theorem and bounds, for which 
these assumptions are relaxed, a violation of the ``similarly 
prepared systems'' condition can mean that the physical situation can no longer be 
adequately described by a contextual Bell model (e.g. due to interaction of the 
particles between ``trials''). Not much can be inferred from apparent violations of Bell 
inequalities in such cases. This point is particularly relevant, 
if a large amount of data is gathered in quick succession. 

We conclude this section with sample data of a 
strongly local, ``noncontextual'' 
Bell model, for which the measured 
arithmetic mean \eqref{eq:BCHSH_statistics_measuredY} violates the bound set 
by the corresponding BCHSH inequality. This shows that statistical violations of 
Bell inequalities can occur, even if they do not occur in the probabilistic model 
itself. In the literature, this phenomenon 
is known as the ``fair sampling loophole''. 

\begin{Example}
	\label{Ex:finite}
\begin{figure}
	\centering
	\includegraphics[width=0.65 \linewidth]{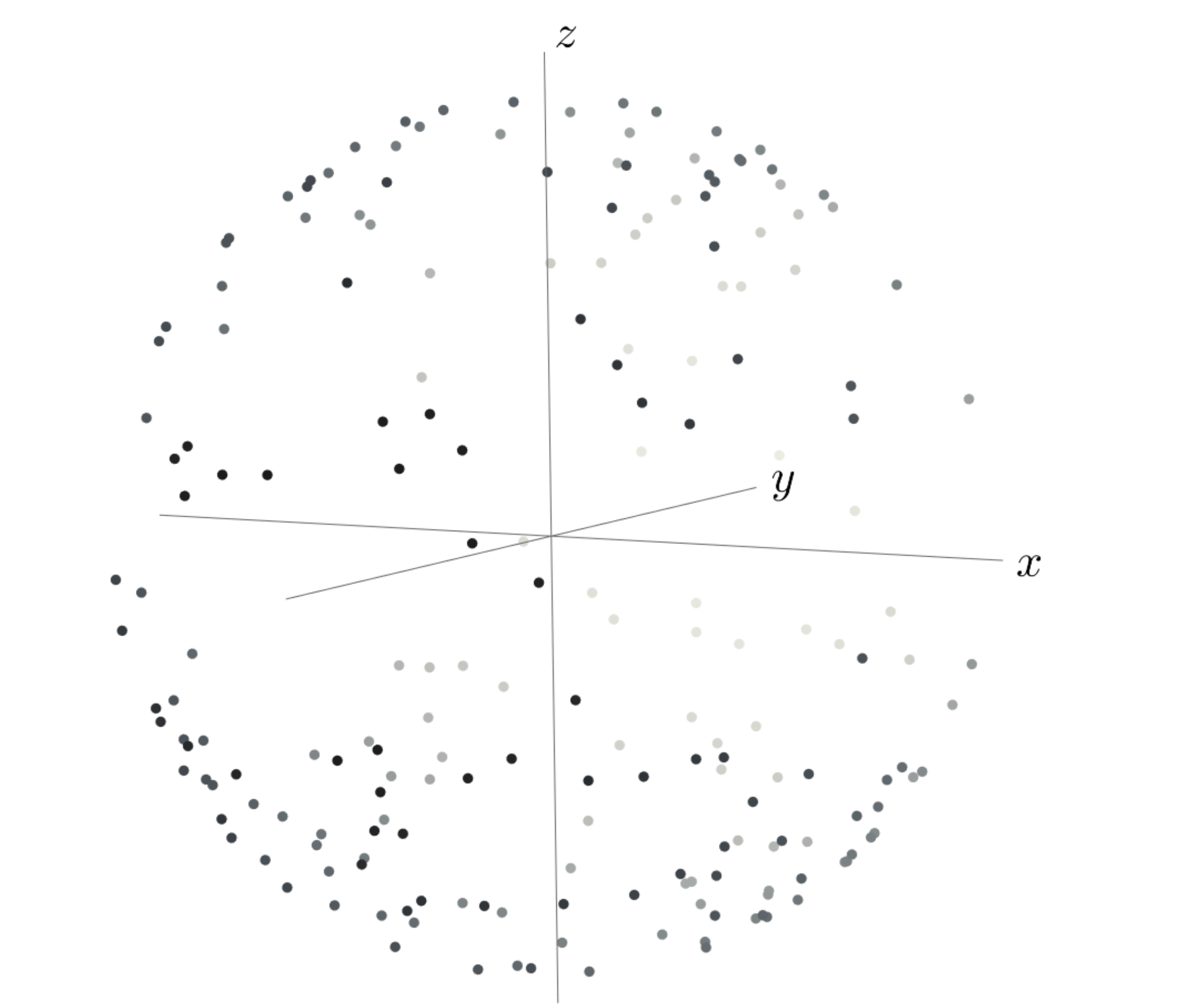}
	\caption{Plot of numerical sample data for the values of $\vec{s}_A$ in the file 
				\texttt{BCHSH\_break.dat}. 
				Data points with larger $y$-values are depicted in a lighter color.}
	\label{fig:data}
\end{figure}
Denote by $\mathcal{N}_{\mu, \sigma^2}$ the (Gaussian) normal distribution 
with expectation 
value $\mu$ and standard deviation $\sigma$ on $\left(\R, \mathcal{B}(\R)\right)$. 
$\operatorname{Rad}_{1/2}$ denotes the Rademacher distribution, 
as in Ex. \ref{Ex:Pab_product}.\ref{itm:Ex:Pab_product2}. 

We construct a probability measure 
$\mathbb{P}$
on the measure space 
\begin{equation}
	\left(\Lambda, \mathcal{A}\right)=\left({\lbrace \pm 1 \rbrace} \times \R^3, 
		2^{\lbrace \pm 1 \rbrace} \otimes \mathcal{B}(\R^3) \right)	
\end{equation}
(cf. Thm. 
14.8 in \cite{klenkeProbabilityTheoryComprehensive2020}) by 
taking the following product: 
\begin{equation}
	\mathbb{P} := \operatorname{Rad}_{1/2} \otimes \, \mathcal{N}_{0,1} 
					\otimes \mathcal{N}_{0,1} \otimes \mathcal{N}_{1,1/2}  \ .
\end{equation}
Thus, any $\lambda \in \Lambda$ may be written as 
$\lambda = \big(\lambda_0, \vec{\lambda} \bigr)$ with 
$\lambda_0 \in \lbrace \pm 1\rbrace$ and $\vec{\lambda} \in \R^3$. $\lambda_0$ 
takes values $\pm 1$ with 
probability $1/2$ each and $\vec{\lambda}$ tends to be ``close'' to the unit 
vector on the third axis $\vec{e}_3$. 

As our unit ``spin vector'' we introduce the random variable 
\begin{equation}
	{s}_A \colon \Lambda \to \mathbb{S}^2 \ , \quad 
		 \big(\lambda_0, \vec{\lambda} \bigr) \mapsto 
	{s}_A \big(\lambda_0, \vec{\lambda} \bigr) 
			= \left(\lambda_1, \lambda_2, \lambda_0 \, \lambda_3\right)
			/{\bigl\vert \vec{\lambda} \bigr\vert} \ .  
\end{equation}
Similarly, we define $\vec{s}_B$ by setting $\vec{s}_B( \lambda):= -\vec{s}_A( \lambda)$ 
for any $\lambda \in \Lambda$. The random variables $A$ and $B$ are chosen  
as in Ex. \ref{Ex:Bell}. We hence obtain a strongly local Bell model, in which 
Bell's correlation assumption \eqref{eq:Bellcorr} holds. 

A simple way to obtain a statistical violation of the BCHSH inequality in this model 
is to generate sample data and to then choose the detector parameters   
appropriately. This is, of course, contrary to experimental 
practice, but it does illustrate the point. 

Numerical sample data for $200$ values of 
$\vec{s}_A$ ($n=50$) along with corresponding computational source code 
is supplied along with this article. The sample vectors are 
illustrated in Fig. \ref{fig:data}. For the data 
set and setting parameters supplied in the supplementary material, we obtain 
the following statistical violation of the BCHSH inequality: 
\begin{equation}
	\bar{Y}_{50}\bigl(\vec{a},\vec{b},\vec{a}',\vec{b}'\bigr)(\boldsymbol{\lambda})	
	\approx 2.24	\ . 
\end{equation}
\end{Example}

\section{Conclusion and outlook} 
\label{sec:conclusion}

We shall summarize the main results. In Sec. \ref{sec:generalizedBell} we found    
that the concept of a Bell model, which 
both Bell's theorem \ref{Thm:Bell} 
and the original BCHSH theorem \ref{Thm:BCHSH} implicitly 
rely on, fails to account for contextuality: 
since in a general Bell experiment the detector settings may influence 
the distribution of the random detector parameters, it is in general not 
admissible to assume the existence of a single setting-independent probability 
measure for all choices of settings. Hence, a general 
probabilistic model for a Bell experiment is needed: 
a contextual Bell model (Def. \ref{Def:generalizedBM}). 
There are two main notions of locality such a model can satisfy: weak and 
strong locality (Def. \ref{Def:generalizedBM_locality}). 
By Lem. \ref{Lem:genBM_strong_implies_weak}, 
strong locality implies weak locality. 
Thm. \ref{Thm:BCHSH_detector} showed that a wide class of strongly local 
contextual Bell models satisfies the BCHSH inequality, generalizing a previous result by 
Gill and Lambare \cite{gillKupczynskisContextualLocally2023}. 
Specifically, those are models for which each detector individually takes the value of the 
``shared variable'' $\lambda_R$ as well as its setting parameter to determine the 
probability for its random detector parameters. 
A physically justified ``classical'' example of such a model 
was given in Exs. \ref{Ex:correlatedP} and \ref{Ex:correlatedP_cont}. 
In Lem. \ref{Lem:uncorrelated+weaklylocal+contextual} we found that 
a correlation between the detector random variables is a necessary 
condition for a violation of the BCHSH inequality, while 
Ex. \ref{Ex:Gumbel} showed that the condition is not sufficient. 
Sec. \ref{sec:poc_example} then provided a ``proof of concept'' that 
contextual Bell models can violate the BCHSH inequality, even if they 
are strongly local. In Sec. \ref{sec:time} we complemented this with a 
nonrigorous discussion of the role of dynamics in Bell experiments, arguing that 
dynamical interactions tend to induce a stochastic dependence between random variables 
associated to the particles as well as the detectors, possibly leading to 
a violation of Bell inequalities in Bell experiments. In Sec. \ref{sec:statistics} 
we addressed 
the general problem of how to test the BCHSH inequality empirically, 
providing both a rigorous approach via the Chernoff-Hoeffding bound 
(Lem. \ref{Lem:Hoeffding}) and a more practical approach via the central limit 
theorem (Prop. \ref{Prop:contextualBellstatistics_clt}). 

Taking a step back, we thus observed that the usual argument against 
a probabilistic description of Bell experiments may be considered 
a ``strawman argument'': a priori one imposes a description via 
``noncontextual'' Bell models, even though this may be inappropriate 
even for ``classical'' 
Bell experiments. Loosely speaking, noncontextuality constitutes a third assumption 
in the derivation of Bell inequalities, which, contrary to the assumptions of 
the existence of 
a probabilistic description and the validity of certain requirements of 
``locality'', is not 
physically natural. And once noncontextuality is dropped as an 
assumption, a Bell inequality does not need to hold unless 
other conditions enforce it. 

With regards to the general debate on the (in)completeness of quantum mechanics 
(cf. Sec. 1), 
it follows that the commonly stated physical consequences of the empirical violation 
of Bell inequalities 
\cite{freedmanExperimentalTestLocal1972,
holtAtomicCascadeExperiments1973,
clauserExperimentalInvestigationPolarization1976,
fryExperimentalTestLocal1976,
aspectExperimentalTestsRealistic1981,
aspectExperimentalRealizationEinsteinPodolskyRosenBohm1982,
aspectExperimentalTestBell1982,
ouViolationBellsInequality1988,
shihNewTypeEinsteinPodolskyRosenBohm1988,
rarityExperimentalViolationBells1990,
kiessEinsteinPodolskyRosenBohmExperimentUsing1993,
kwiatNewHighIntensitySource1995,
tittelViolationBellInequalities1998,
weihsViolationBellInequality1998,
panExperimentalTestQuantum2000,
roweExperimentalViolationBell2001,
matsukevichEntanglementRemoteAtomic2006,
ursinEntanglementbasedQuantumCommunication2007,
matsukevichBellInequalityViolation2008,
ansmannViolationBellsInequality2009,
hofmannHeraldedEntanglementWidely2012,
vermeydenExperimentalViolationThree2013,
giustinaBellViolationUsing2013,
christensenDetectionLoopholeFreeTestQuantum2013,
hensenLoopholefreeBellInequality2015,
giustinaSignificantLoopholeFreeTestBell2015,
shalmStrongLoopholeFreeTest2015,
rosenfeldEventReadyBellTest2017,
yinSatellitebasedEntanglementDistribution2017} 
are not based on theoretically firm arguments. 
In conjunction with the circular argument implicit in other known 
``no-go theorems'', this provides further incentive to seek 
descriptions of quantum phenomena via mathematical 
probability theory. 

Indeed, in part I \cite{reddigerApplicabilityKolmogorovsTheory2025,
reddigerAddendumApplicabilityKolmogorovs2026} it was argued that there is 
a natural approach to such a description. It was further shown that this 
approach, geometric quantum theory, is in principle able to make novel predictions, 
so that the problem may be  
moved from the realm of metaphysical ``no-go theorems'' to that of direct  
experimental comparison. For this purpose, part III of this series provides 
a general discussion of measurement in geometric quantum theory with a particular 
focus on the quantum-mechanical projection postulate. 

Depending on how the question of the applicability of mathematical 
probability to the description of quantum phenomena is resolved, there will also be 
consequences to the theory of 
``quantum statistics'' \cite{petzQuantumInformationTheory2008}. 
The standard theory of statistics \cite{schervishTheoryStatistics1997} 
is based on mathematical probability theory, so that a resolution of the 
question in favor of the latter would also undermine the raison d'être of 
the theory of ``quantum statistics''. 

Future work should first and foremost focus on the construction of 
an example of a strongly local contextual Bell model that is based on 
physical dynamics and that violates the BCHSH inequality for 
certain settings. The 
challenge is to go beyond the mere mathematical example in Sec. \ref{sec:poc_example}, 
using either classical or quantum dynamics. Since it was argued in 
Sec. \ref{sec:time} that such a model is only of interest, if it is relativistic, 
a classical, special-relativistic model seems to be the appropriate next step. 
The use of classical electromagnetism for this purpose with 
stochastically dependent initial positions 
of the particles suggests itself. In principle one may also introduce a 
stochastic dependence between some initial higher order electromagnetic multipole moments.
In the search for contextual models based on 
quantum dynamics, the detector model in \cite{reddigerSolutionQuantumTime2026} 
may be of use, if generalized to the special-relativistic bipartite setting. 
Following up on Rem. \ref{Rem:reldynamics}.\ref{itm:Rem:reldynamics1}, it 
is a general theoretical question for models based on (relativistic) 
quantum dynamics, whether there exist examples in which strong locality is violated yet 
weak locality is not \cite{gisinCanRelativityBe2015}. 

Provided that dynamical examples in the aforementioned sense can be found, 
one may ask whether the theory of contextual Bell models has applications 
beyond the objectives of this work. To date, Bell's work has inspired  
contributions to ``quantum information theory'' 
\cite{lambropoulosFundamentalsQuantumOptics2007,
nielsenQuantumComputationQuantum2016} and the subject of 
``quantum communication'' \cite{keylFundamentalsQuantumInformation2002,
nobelcommitteeforphysicsScientificBackgroundNobel2022}. 
It is natural to explore consequences for these 
subject areas. 

On a final note, it seems that the conflict between the groups of scientists 
arguing for and against the completeness of quantum mechanics 
 in the history of 
``no-go theorems'' 
is a subject of potential interest to the sociology of science. 
At present, this historical conflict is well-documented \cite{
pinchWhatDoesProof1977,
pinchHiddenvariablesControversyQuantum1979,
cushingQuantumMechanicsHistorical1994,
bellerQuantumDialogueMaking1999,
freireStoryEndingQuantum2003,
freirejuniorQuantumDissidentsRebuilding2015,
dieksNeumannImpossibilityProof2017,
bacciagaluppiEinsteinParadoxDebate2024}. What are the driving factors that, at 
least among some scientists, seem to 
turn a genuine subject of scientific inquiry into an ideological 
debate? Why are, as it appears and contrary to  
good scientific practice, counterhypotheses not taken seriously (enough)? Part of the 
answer may lie in the perception, that metaphysical questions are 
at stake. The competition for positions, funding, and academic prestige 
between research communities 
aiming to answer similar questions 
with contending research programs may be another ``piece of the puzzle''. 
Still, Bell himself was an example, that the dividing line between the two groups  
need not be sharp, for he discovered an important ``no-go theorem'' 
while still believing in a 
deeper description of quantum phenomena \cite{bellImpossiblePilotWave1982,
bellSpeakableUnspeakableQuantum2004,
merminHiddenVariablesTwo1993}. Rather than employing ``no-go theorems'' as a means 
to suppress research in this direction, Bell viewed them 
``as identifying conditions that such a description would
have to meet'' \cite{merminHiddenVariablesTwo1993}. 

While the results of this work do not support 
``closing the door'' on the completeness debate  
\cite{aspectClosingDoorEinstein2015}, they may serve to remind us researchers in 
the foundations of quantum theory of the following quote by Einstein: ``Subtle is the Lord, but malicious He is not.'' 

\section*{Acknowledgements}	

Wolfgang Paul and Kenneth Wharton deserve acknowledgement for helpful discussion. 
Furthermore, the author would like to thank Igor Volobouev for an introduction to Bell's original result, Claudio Emmrich for incentivizing a deeper study of the subject, and Dzmitry Matsukevich for providing empirical sample data. 

%

\section*{Data availability}


Supplementary source code for Ex. \ref{Ex:finite} is provided along with this article. 

\section*{Disclosure statement}	

The author does not have any conflict of interest to declare.

\printbibliography

\end{document}